\documentclass[showpacs,twocolumn]{revtex4}

\usepackage{graphicx}
\usepackage{dcolumn}
\usepackage{amsmath}
\usepackage{color}

\makeatletter
\def\btt#1{\texttt{\@backslashchar#1}}
\DeclareRobustCommand\bblash{\btt{\@backslashchar}} \makeatother

\begin{document}
\title{Cloud of strings in third-order Lovelock gravity}
\author{Sushant G. Ghosh $^{a \;b}$} \email{sghosh2@jmi.ac.in, sgghosh@gmail.com}
\author{Uma Papnoi $^{b}$}
\author{Sunil D. Maharaj $^{a}$}\email{maharaj@ukzn.ac.za}
\affiliation{$^{a}$ Astrophysics and Cosmology
Research Unit, School of Mathematics, Statistics and Computer Science, University of
KwaZulu-Natal, Private Bag 54001, Durban 4000, South Africa}
\affiliation{$^{b}$ Centre for Theoretical Physics, Jamia Millia
Islamia, New Delhi 110025, India}
\date{\today}
\begin{abstract}
Lovelock theory is a natural extension of the Einstein theory of general relativity to higher dimensions in which the first and second orders correspond, respectively, to  general relativity and Einstein-Gauss-Bonnet gravity.  We present exact black hole solutions of $D\geq 4$-dimensional spacetime for first-, second-, and third-order Lovelock gravities in a string cloud background. Further, we compute the mass, temperature, and entropy of black hole solutions for the higher-dimensional general relativity and Einstein-Gauss-Bonnet theories and also perform thermodynamic stability of black holes. It turns out that the presence of the  Gauss-Bonnet term and/or background string cloud completely changes the black hole thermodynamics. Interestingly, the entropy of a black hole is unaffected due to a background string cloud. We rediscover several known spherically symmetric black hole solutions in the appropriate limits.

\end{abstract}

\maketitle

\section{\bigskip INTRODUCTION}
Lovelock gravity is one of the natural generalization of Einstein's general relativity, introduced by David Lovelock \cite{dll}, the action of which contains higher-order curvature terms. The Lovelock action in $D (\geq
4)$-dimensional spacetime reads
\begin{align}
\label{action}
I=& \frac{1}{2\kappa_D^2}\int  d^D x\sqrt{-g}\sum_{p=0}^{[D/2]}\alpha_{(p)}{\mathcal L}_{(p)}+I_{ matter},\\
 {\mathcal L}_{(p)}:=&\frac{1}{2^p}\delta^{\mu_1\cdots \mu_p\nu_1\cdots \nu_p}_{\rho_1\cdots \rho_p\sigma_1\cdots \sigma_p}R_{\mu_1\nu_1}^{\phantom{\mu_1}\phantom{\nu_1}\rho_1\sigma_1}\cdots R_{\mu_p\nu_p}^{\phantom{\mu_p}\phantom{\nu_p}\rho_p\sigma_p},
\end{align}
where $\kappa_D$ is a constant related to $G_D$ via $\kappa_D := \sqrt{8\pi G_D}$ with $\kappa_D^2>0$, the coupling constant $\alpha_{(p)}$ has dimension of $({\rm length})^{2(p-1)}$, and $\mathcal{L}_{(p)}$ is  the Euler density of a 2$p$-dimensional
manifold. The symbol $\delta$ describes a totally antisymmetric
product of Kronecker deltas, normalized to take values $0$ and $\pm
1$~\cite{dll,km2006}, defined by
\begin{align}
\delta^{\mu_1\cdots \mu_p}_{\rho_1\cdots \rho_p}:=&p!\; \delta^{\mu_1}_{[\rho_1}\cdots \delta^{\mu_p}_{\rho_p]},
\end{align}
where $\alpha_{(0)}$ is related to the cosmological constant $\Lambda$ by $\alpha_{(0)}=-2\Lambda$. The Lovelock action $I$ reduces to the Einstein-Hilbert action in four dimensions, and its
second term is the Gauss-Bonnet invariant.
Lovelock theories are distinct, among the  larger class of general higher-curvature theories, in having
field equations involving not more than second derivatives of the metric. Consequently,
Lovelock gravity theories are free from many of the pathologies that affect general higher
derivative gravity theories.

As higher-dimensional members of Einstein's general relativity
family, Lovelock gravities allow us to explore several conceptual
issues of gravity in depth in a broader setup. Most interestingly,
one can include features of black holes such as their existence and
uniqueness theorems, their thermodynamics, the definitions of their
mass and entropy, their horizon properties, etc.  Also, such a
theory may be used in the context of the AdS/CFT correspondence to
investigate the effects of including higher-curvature terms
\cite{fms}.    It has, therefore, been explored to a large extent,
also possibly for its appearance in strings theories at low energies
\cite{Gross}.  In this paper, we will be concerned with the black
hole solutions of this Lovelock theory, and we will discuss how
higher-curvature corrections to black hole physics substantially
change the qualitative features we know from our experience with
black holes in general relativity. Since their inception, steady
attention has been devoted to black hole solutions, including their
formation, stability, and thermodynamics.  The spherically symmetric
static black hole  solution for second-order Lovelock gravity (the
theory that is usually referred as the Einstein-Gauss-Bonnet theory)
was  first  obtained by Boulware and Deser \cite{bd}, and this kind of
solution for third-order Lovelock gravity was introduced in Ref. \cite{ds}. 
Exact black hole solutions of the former can be found in Ref. \cite{egb}
and the latter in Refs. \cite{ll,som,som1}. The black hole solution in 
Einstein-Gauss-Bonnet theory in a string cloud model was considered in Refs.  \cite{hr,som1}.

The recent theoretical developments signal toward a scenario in
which the fundamental building blocks of the Universe are extended
objects instead of point objects and have been considered quite
seriously \cite{synge}.  The most natural and popular candidate is
one-dimensional strings object. This resulted in the intense
level of activity towards the study of the gravitational effects of matter in
the form of clouds of both cosmic and fundamental strings \cite{Le1,
sgr,hr,sgsm}. In addition, the intense level of activity in string
theory has led to the idea that many of the classic vacuum schemes,
such as the static Schwarzschild black hole (point mass), may have
atmospheres composed of a fluid or field of strings. Further, this
two-fluid atmosphere can model an array of physical situations at
diverse distance scales, which  can depict the atmosphere around a
black hole with a distance scale of multiples of Schwarzschild
radii. It could also describe a globular cluster with components of
a dark matter at a scale tuned to the order of parsecs. The event
horizon for the classical Schwarzschild metric in the background of
a cloud of strings  has a modified radius $ r_H={2M}/{(1-a)}$ with $a$
as a string cloud parameter \cite{Le1}, thereby enlarging the
Schwarzschild radius of the black hole by the factor $(1-a)^{-1}$,
which may have several astrophysical consequences, e.g., on a wormhole
\cite{mrcs}. Further, Glass and Krisch \cite{gk} have demonstrated
that allowing the Schwarzschild mass  to be a function of radial
position builds an atmosphere with a string fluid stress energy
around a static, spherically symmetric, object. Thus, the study of
Einstein's equations coupled with a string cloud, in both general
relativity and modified theories, may be very important because
relativistic strings at a classical level can be used to construct
applicable models \cite{synge}.

Intense activity of studying black hole solutions in modified
theories of gravity including Lovelock theories of gravity is due
to the fact that, besides theoretical results, cosmological evidence,
e.g., dark matter and dark energy, the possibility of
modification of the Einstein gravity is suggested. Many other authors have found
exact black hole solutions with a string cloud background, for
instance, in general relativity \cite{Le1}, in Einstein-Gauss-Bonnet
gravity \cite{hr,som1}, and also in Lovelock gravity \cite{sgr,sgsm}.

It is the purpose of this paper to obtain an exact black hole
solution in second- and third-order Lovelock theories with a cloud
of strings in the background. We shall present a class of black hole
solutions  endowed with a cloud of strings. In particular, we explicitly
 bring out how the effect of the background string cloud can alter
 black hole solutions and their properties as we know from our
knowledge of  black holes in general relativity. Our
attention will be given to the second-order Einstein-Gauss-Bonnet case, which
exhibits most of the relevant qualitative features. We obtain
$D$-dimensional static spherically symmetric black hole solutions in
a string cloud background with the three terms of Lovelock gravity
that are Einstein or general relativity, Gauss-Bonnet and third-order
 Lovelock terms. We analyze their thermodynamic properties and
also perform a stability analysis the for Einstein-Gauss-Bonnet
case.

\section{EINSTEIN LOVELOCK ACTION}
Lovelock gravity is the most general second-order gravity theory
in higher-dimensional spacetimes, which is free of ghosts \cite{dll}. 
The Lovelock tensor is nonlinear in the Riemann tensor and nontrivially differs from the Einstein tensor
only if $D \geq 4$. The third-order Lovelock action  with matter in $D \geq 4$
dimensions reads
\cite{dll}%
\begin{equation}
\mathcal{I}_{G}=\frac{1}{2}\int_{\mathcal{M}}dx^{D}\sqrt{-g}\left[  \mathcal{L}_{1} +\alpha_{2}\mathcal{L}_{GB}+\alpha_{3}\mathcal{L}_{\left(  3\right)
} \right] + \mathcal{I}_{S}.
\end{equation}
with $\kappa_D=1$. The Einstein term $ \mathcal{L}_1 = R$, the second-order 
Lovelock (Gauss-Bonnet) term $\mathcal{L}_{GB}$  is
\begin{equation}
\mathcal{L}_{GB}=R_{\mu\nu\gamma\delta}R^{\mu
\nu\gamma\delta}-4R_{\mu\nu}R^{\mu\nu}+R^{2},
\end{equation} and
\begin{align}
\mathcal{L}_{\left(  3\right)  }  &  =2R^{\mu\nu\sigma\kappa}R_{\sigma
\kappa\rho\tau}R_{\quad\mu\nu}^{\rho\tau}+8R_{\quad\sigma\rho}^{\mu\nu
}R_{\quad\nu\tau}^{\sigma\kappa}R_{\quad\mu\kappa}^{\rho\tau}\nonumber\\
&  +24R^{\mu\nu\sigma\kappa}R_{\sigma\kappa\nu\rho}R_{\ \mu}^{\rho}%
+3RR^{\mu\nu\sigma\kappa}R_{\sigma\kappa\mu\nu}\nonumber\\
&  +24R^{\mu\nu\sigma\kappa}R_{\sigma\mu}R_{\kappa\nu}+16R^{\mu\nu}%
R_{\nu\sigma}R_{\ \mu}^{\sigma}\\
&  -12RR^{\mu\nu}R_{\mu\nu}+R^{3},\nonumber
\end{align}
is the third-order Lovelock Lagrangian. Here, $R_{\mu\nu}$, $R_{\mu\nu\gamma\delta\text{
}}$, and $R$  are the Ricci tensors, Riemann tensors, and  Ricci scalar, respectively. 
The variation of the action with respect to the metric $g_{\mu\nu}$ gives the 
Einstein-Gauss-Bonnet-Lovelock equations,%
\begin{equation}\label{ee}
G_{\mu\nu}^{E}+\alpha_{2}G_{\mu\nu}^{GB}+\alpha_{3}G_{\mu\nu}^{\left(
3\right)  }=T_{\mu\nu}^{S},
\end{equation}
where
$G_{\mu\nu}^{E}$ is the Einstein tensor, while $G_{\mu\nu}^{GB}$ and
$G_{\mu\nu}^{\left(  3\right)  }$ are given explicitly, respectively, by \cite{km2006}%
\begin{eqnarray}
 G_{\mu\nu}^{GB} & = & 2\;\Big( -R_{\mu\sigma\kappa\tau}R_{\quad\nu}^{\kappa
\tau\sigma}-2R_{\mu\rho\nu\sigma}R^{\rho\sigma}-2R_{\mu\sigma}R_{\ \nu
}^{\sigma} \nonumber \\ & &  +RR_{\mu\nu}\Big) -\frac{1}{2}\mathcal{L} _{GB}g_{\mu\nu},
\end{eqnarray}
and
\begin{gather}
G_{\mu\nu}^{\left(  3\right)  }=-3\left(  4R_{\qquad}^{\tau\rho\sigma\kappa
}R_{\sigma\kappa\lambda\rho}R_{~\nu\tau\mu}^{\lambda}-8R_{\quad\lambda\sigma
}^{\tau\rho}R_{\quad\tau\mu}^{\sigma\kappa}R_{~\nu\rho\kappa}^{\lambda}\right. \nonumber
\\
+2R_{\nu}^{\ \tau\sigma\kappa}R_{\sigma\kappa\lambda\rho}R_{\quad\tau\mu
}^{\lambda\rho}-R_{\qquad}^{\tau\rho\sigma\kappa}R_{\sigma\kappa\tau\rho
}R_{\nu\mu}+8R_{\ \nu\sigma\rho}^{\tau}R_{\quad\tau\mu}^{\sigma\kappa
}R_{\ \kappa}^{\rho}\nonumber\\
+8R_{\ \nu\tau\kappa}^{\sigma}R_{\quad\sigma\mu}^{\tau\rho}R_{\ \rho}^{\kappa
}+4R_{\nu}^{\ \tau\sigma\kappa}R_{\sigma\kappa\mu\rho}R_{\ \tau}^{\rho
}-4R_{\nu}^{\ \tau\sigma\kappa}R_{\sigma\kappa\tau\rho}R_{\ \mu}^{\rho
}\nonumber\\
+4R_{\qquad}^{\tau\rho\sigma\kappa}R_{\sigma\kappa\tau\mu}R_{\nu\rho}%
+2RR_{\nu}^{\ \kappa\tau\rho}R_{\tau\rho\kappa\mu}+8R_{\ \nu\mu\rho}^{\tau
}R_{\ \sigma}^{\rho}R_{\ \tau}^{\rho}\nonumber\\
-8R_{\ \nu\tau\rho}^{\sigma}R_{\ \sigma}^{\tau}R_{\ \mu}^{\rho}-8R_{\quad
\sigma\mu}^{\tau\rho}R_{\ \tau}^{\sigma}R_{\nu\rho}-4RR_{\ \nu\mu\rho}^{\tau
}R_{\ \tau}^{\rho}\nonumber\\
+4R_{\quad}^{\tau\rho}R_{\rho\tau}R_{\nu\mu}-8R_{\ \nu}^{\tau}R_{\tau\rho
}R_{\ \mu}^{\rho}+4RR_{\nu\rho}R_{\ \mu}^{\rho}
\left.  -R^{2}R_{\nu\mu}\right) \nonumber \\ -\frac{1}{2}\mathcal{L}_{\left(  3\right)
}g_{\mu\nu},\nonumber
\end{gather}
and $T_{\mu\nu}$ is the energy-momentum tensor of matter that we consider as
 a cloud of strings.  Note that for third-order Lovelock gravity, the
nontrivial third term requires that the dimension of spacetime should satisfy  $D \geq 7 $.
\section{STRING-CLOUD MODEL}
Let us consider a cloud of strings as matter. For completeness,
we give a brief review of the theory of a cloud of strings (see Ref. \cite{Le1} for further details).
The Nambu-Goto action of a string evolving in   spacetime is given by
\begin{equation}    I_{\mathcal{S}} = \int_{\Sigma} \mathcal{L} \;  d\lambda^{0} d\lambda^{1}, \hspace{0.2in} \mathcal{L} = m (\gamma)^{-1/2},\end{equation}
where $m$ is a positive constant, $\lambda^{0}$ and $\lambda^{1}$ being timelike and  spacelike parameters \cite{synge}.  The string world sheet $\Sigma$ is given by
\begin{equation}
\gamma_{a b} = g_{\mu \nu} \frac{\partial x^{\mu}}{\partial \lambda^{a}} \frac{\partial x^{\nu}}{\partial \lambda^{b}},
\end{equation}
and $\gamma $ = det $\gamma_{a b}$. Associated with the strings world sheet, we have the bivector of the form
\begin{equation}
\label{eq:bivector}
\Sigma^{\mu \nu} = \epsilon^{a b} \frac{\partial x^{\mu}}{\partial \lambda^{a}} \frac{\partial x^{\nu}}{\partial \lambda^{b}},
\end{equation}where $\epsilon^{a b}$ denotes the two-dimensional Levi-Civit$\acute{a}$ tensor given by $\epsilon^{0 1} = - \epsilon^{1 0} = 1$.
Within this setup, the Lagrangian density becomes
\[\mathcal{L} = m \left[-\frac{1}{2} \Sigma^{\mu \nu} \Sigma_{\mu \nu}\right]^{1/2}.     \]
Further, since $T^{\mu \nu} = 2 \partial \mathcal{L}/\partial g^{\mu \nu}$, we obtain the energy-momentum tensor for one string as
\begin{equation}
T^{\mu \nu} = m \Sigma^{\mu \rho} \Sigma_{\rho}^{\phantom{\rho} \nu}/(-\gamma)^{1/2}.
\end{equation}
Hence, the energy-momentum tensor for a cloud of strings is
\begin{equation}
T^{\mu \nu} = \rho {\Sigma^{\mu \sigma} \Sigma_{\sigma}^{\phantom{\sigma} \nu}}/{(-\gamma)^{1/2}  },
\end{equation}where $\rho$ is the proper density of a string cloud and quantity $\rho \; (\gamma)^{-1/2} $ is the gauge-invariant density. The strings is characterized by a surface-forming bivector $\Sigma^{\mu \nu} $ and conditions to be a surface-forming are
\begin{eqnarray} \label{eq:imp}
& & \Sigma^{\mu [\alpha} \Sigma^{\beta \gamma]} = 0, \nonumber \\ & & \nabla_{\mu} \Sigma^{\mu [\alpha} \Sigma^{\beta \gamma]} = 0,
\end{eqnarray}
where the square brackets in Eq.~(\ref{eq:imp}) indicate antisymmetrization. The above equation, in conjunction with Eq.~(\ref{eq:bivector}), leads to the useful identity
\begin{equation}
\label{eq:usable}
\Sigma^{\mu \sigma} \Sigma_{\sigma \tau} \Sigma^{\tau \nu} = \gamma \Sigma^{\nu \mu},
\end{equation}
which will be used in subsequent calculations. Further, the conservation of energy-momentum tensor $T^{\mu \nu}_{;\nu} =0$ implies that
\begin{equation}
\nabla_{\mu} \left(\rho \Sigma^{\mu \sigma}\right) \Sigma_{\sigma}^{\phantom{\sigma} \nu}/{(-\gamma)^{1/2}}
+ \rho \Sigma^{\mu \sigma} \nabla_{\mu}  \left(\Sigma_{\sigma}^{\phantom{\sigma} \nu}/{(-\gamma)^{1/2}  }\right) =0,
\end{equation}
which upon multiplication by $\Sigma_{\nu \alpha}/{(-\gamma)^{1/2}  }$ leads to
$\nabla_{\mu} (\rho \Sigma^{\mu \sigma}) \Sigma_{\sigma}^{\phantom{\sigma} \nu} \Sigma_{\nu \alpha}/\gamma = 0$.
Contracting the previous identity with $\Sigma_{\alpha \nu}$ and using  Eq.~(\ref{eq:usable}), we obtain
$\nabla_{\mu} (\rho \Sigma^{\mu \sigma}) \Sigma_{\sigma}^{\phantom{\sigma} \nu} = 0$,
and finally adapting to parametrization, we get
\begin{equation}
\label{eq:div}
\partial_{\mu} (\sqrt{-\mathbf{g}} \rho \Sigma^{\mu \sigma}) = 0.
\end{equation}

\section{Field equations}
Here, we want to obtain $D$-dimensional static spherically symmetric solutions of Eq.~(\ref{ee}) with a cloud of strings as source and investigate its properties. We assume that the metric has the form
\begin{equation}\label{metric}
ds^2 = -f(r) dt^2+ \frac{1}{f(r)} dr^2 + r^2 \tilde{\gamma}_{ij}\; dx^i\; dx^j,
\end{equation}
where $ \tilde{\gamma}_{ij} $ is the metric of a $(D-2)$-dimensional constant curvature space $k = 1,\; 0,\;$ or -1. In this paper, we shall stick to $k = 1$.  To find the metric function $f(r)$, we consider the
components of Eq.~(\ref{ee}).  Using this metric \textit{ansatz}, the Einstein-Gauss-Bonnet-Lovelock $rr$  equation of motion  (\ref{ee}) reduces to 
\begin{eqnarray}\label{master}
& & \left[r^{5}-2\tilde{\alpha}_{2}r^{3}\left(  f\left(  r\right)  -1\right)
+3\tilde{\alpha}_{3}r\left(  f\left(  r\right)  -1\right)  ^{2}\right]
f^{\prime}\left(  r\right)  +\nonumber\\
& & \left(  n-1\right)  r^{4}\left(  f\left(  r\right)  -1\right)  -\left(
n-3\right)  \tilde{\alpha}_{2}r^{2}\left(  f\left(  r\right)  -1\right)
^{2}+\nonumber\\
& & \left(  n-5\right)  \tilde{\alpha}_{3}\left(  f\left(  r\right)  -1\right)^{3} = \frac{2r^6 }{n}T^r_r,
\end{eqnarray}
in which a prime denotes a derivative with respect to $r,$ $n=D-2,$
$\tilde{\alpha}_{2}=$ $\left(  n-1\right)  \left(  n-2\right)  \alpha_{2}$, and
$\tilde{\alpha}_{3}=$ $\left(  n-1\right)  \left(  n-2\right)  \left(
n-3\right)  \left(  n-4\right)  \alpha_{3}$.
In general, Eq.~(\ref{master}) has one real and two complex solutions. It may have three real solutions as well under some conditions. Here, we consider only the real solution.

Here, the density $\rho$ and the bivector $\Sigma_{\mu\nu}$ are the functions of $r$ only as we  seek static spherically symmetric solutions. The only surviving component of the bivector $\Sigma$ is $\Sigma^{tr} =  - \Sigma^{rt}$.  Thus, $T^t_t=T^r_r=-\rho \Sigma^{tr}$, and  from Eq.~(\ref{eq:div}), we obtain  $\partial_{r} (\sqrt{r^n T^t_t}) = 0$, which implies
\begin{equation}
T^t_t = T^r_r = \frac{a}{r^n},
\end{equation}
for some real constant $a$.
In the rest frame associated with the observer, the energy density of the matter
will be given by $\rho = {a}/{r^n}.$
The energy-momentum of the source can be written as
\begin{equation} T^{\mu}_{\nu} =
\frac{a}{r^{n}} \mbox{{{diag}}}[1, 1, 0,\ldots, 0]. \label{emt}
\end{equation}
The weak energy condition demands that for every timelike vector field -the matter density observed by the corresponding observer is always non-negative:
\begin{equation}
\rho \geq 0,\hspace{0.1 in}\rho+ P_{r} \geq 0. \label{wec}
\end{equation}
 The strong energy condition requires
\begin{equation}
\rho + P_{r} \geq 0, \rho + n P_{i} \geq 0. \hspace{0.1 in}
\end{equation}
Clearly, all energy conditions hold.

In three and four dimensions, Lovelock theory coincides with Einstein theory \cite{xc}, e.g., for $D=4\left(  \text{i.e.,
}n=2\right)$, we get%
\begin{equation}
r^{3}f^{\prime}\left(  r\right)  +  f\left(  r\right)  -1=a,
\end{equation}
which clearly is independent of  $\alpha_{2}$ and $\alpha_{3}$,  and therefore it will be the
Einstein equation in four dimensions admitting the solution
\begin{equation}\label{lfour dimensions}
f\left(  r\right)  =1-\frac{2m}{r}+ a.
\end{equation}
This solution was first obtained by Letelier \cite{Le1}, and the metric represents the black hole spacetime associated with a spherical mass $m$ centered at the origin of the system of coordinates, surrounded by a spherical cloud of strings. The event horizon of the metric is placed at $r_{EH}=2m/(1-a)$.  In the limit $a \rightarrow 0$, we recover the Schwarzschild radius, and close to unity. the event horizon radius tends to infinity. On the other hand, a cloud of strings alone ($m = 0$) does not  have a horizon; it represents only a naked singularity at $r = 0$. Besides, the metric (\ref{metric}) with (\ref{lfour dimensions}) can be understood as the metric associated with a global monopole.

But in higher dimensions, the Lovelock theories  are actually different. In fact, for $D > 4$, Einstein gravity can be thought of as a particular case of Lovelock gravity since the Einstein-Hilbert term is one of several terms that constitute the Lovelock action. Hence, for  $D > 4$ and $\alpha_{2} = \alpha_{3} = 0 $, we obtain
\begin{equation}\label{eehd}
f(r)=     {r}^{5}f'(r) + \left( n-3 \right) {r}^{4} \left( f \left( r
\right) -1 \right)   = \frac {2a}{n{r}^{n-6}},
\end{equation}
which admits the solution
\begin{equation}\label{ndsol}
f\left(  r\right)  =1-\frac{2m}{(n-1)r^{n-1}}- \frac{2a}{nr^{n-2}}.
\end{equation}
We also observe that the four-dimensional solution (\ref{lfour dimensions}) is recovered in the limit $n \rightarrow 2$.  The integration constant $m$ in Eq.~(\ref{ndsol}) is related to the Arnowitt-Deser-Misner (ADM) mass $M$ via
\begin{equation}\label{a2}
m =\frac{16 \pi M}{n V_{n}},\;\;\;  V_n=\frac{2\pi^{(n+1)/2}}{\Gamma{(n+1)/2}},
\end{equation}  where $V_n$ is the volume of the $(D-2)$-dimensional unit sphere.

To study the general structure of the solution given by (\ref{eehd}),  we look for the essential singularity.  It is seen that the Kretschmann scalar ($\mathcal{K} =
R_{abcd} R^{abcd}$) for the metric (\ref{metric}) reduces to
\begin{eqnarray}
\mathcal{K}= R_{abcd}\;R^{abcd}  = f{''}^2(r) +2n\frac{f'(r)}{r^2} + 2 n (n-1)  \frac{f(r)^2}{r^4} ,  \label{density} \nonumber
\end{eqnarray}which on inserting (\ref{eehd}) becomes
\begin{eqnarray}\label{ks}
\mathcal{K}&=& \frac{4n^2(n^2-2)m^2}{r^{2n+2}} + \frac{8n(n-1)^2am}{r^{2n+1}} + \frac{4a^2}{n^2 r^{2n}} (7n^2 \nonumber \\ && +n^4-4n^3-6n+4).
\end{eqnarray}
The Kretschmann scalar (\ref{ks}) diverges as $r \rightarrow 0$, indicating the scalar polynomial or essential singularity at $r=0$.  It is interesting to see that the metric (\ref{metric}) is well behaved even if $m=0$, as (\ref{ks}) indicates.

In the higher-dimensional case, a fact that deserves to be emphasized is that a cloud of strings alone, unlike in four dimensions, can have an event horizon located at $r_{SEH}= (2a/n)^{1/(n-2)}$. Thus, we have extended the Letelier \cite{Le1} solutions to higher-dimensional spacetime. Now, we look for the existence of event horizons, and therefore for possible black hole solutions. The horizons, if they exist, are given by zeros of $f(r)=g^{rr}=0$. The black hole horizon of the solution (\ref{eehd}) is located at, e.g., in the five-dimensional case, $r_{EH}=a\pm \sqrt{{a^2+18m}}/3$ and at $r=r_{EH} = \eta^{1/3}/6+a/ \eta^{1/3}$ with $\eta:=216m + 6 \sqrt{1296m^2-6a^3}$ for six-dimensional case.

We note that the gravitational mass of a black hole is determined by $f(r_+)=0$, which, from Equation~(\ref{ndsol}), reads
\begin{eqnarray}
M = \frac{n (n-1) V_n}{32\pi}r_{+}^{n-1} \left( 1 - \frac{2 a}{n r_{+}^{n-2}}\right). \label{mass}
\end{eqnarray}Equation~(\ref{mass}), takes the form of the $D$-dimensional Schwarzschild black hole when $a\rightarrow0$.
\begin{figure*}
\begin{tabular}{|c|c|c|}
\hline
\includegraphics[width= 8 cm, height= 6 cm]{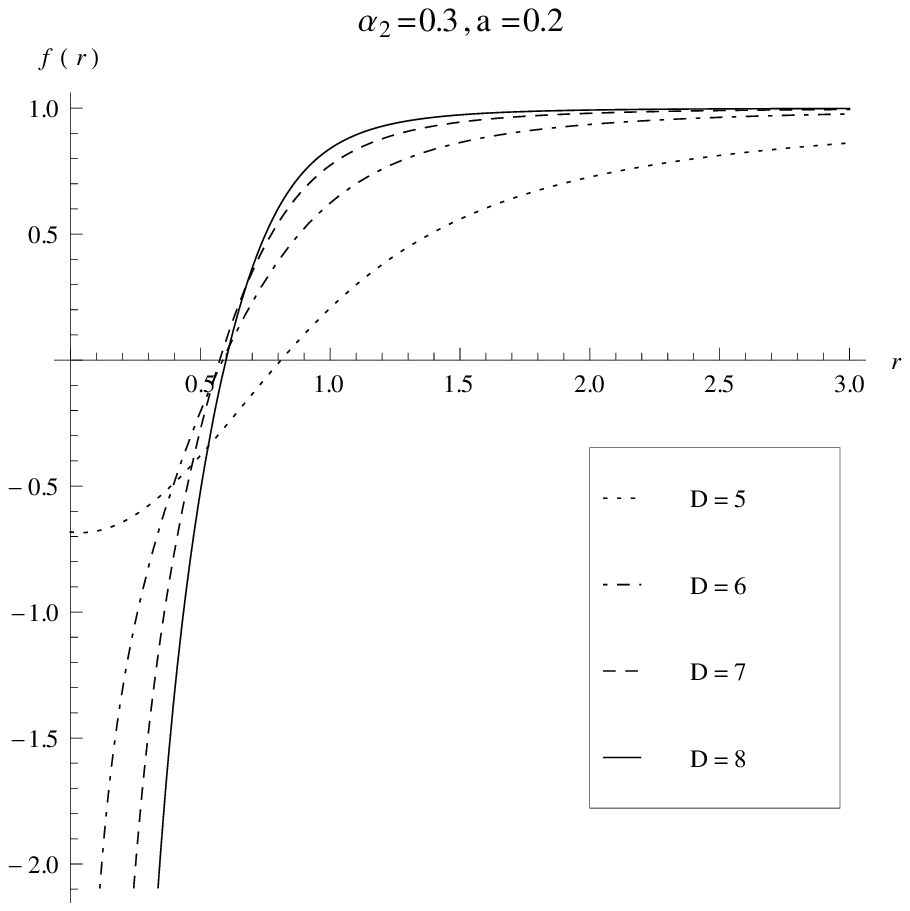}&
\includegraphics[width= 8 cm, height= 6 cm]{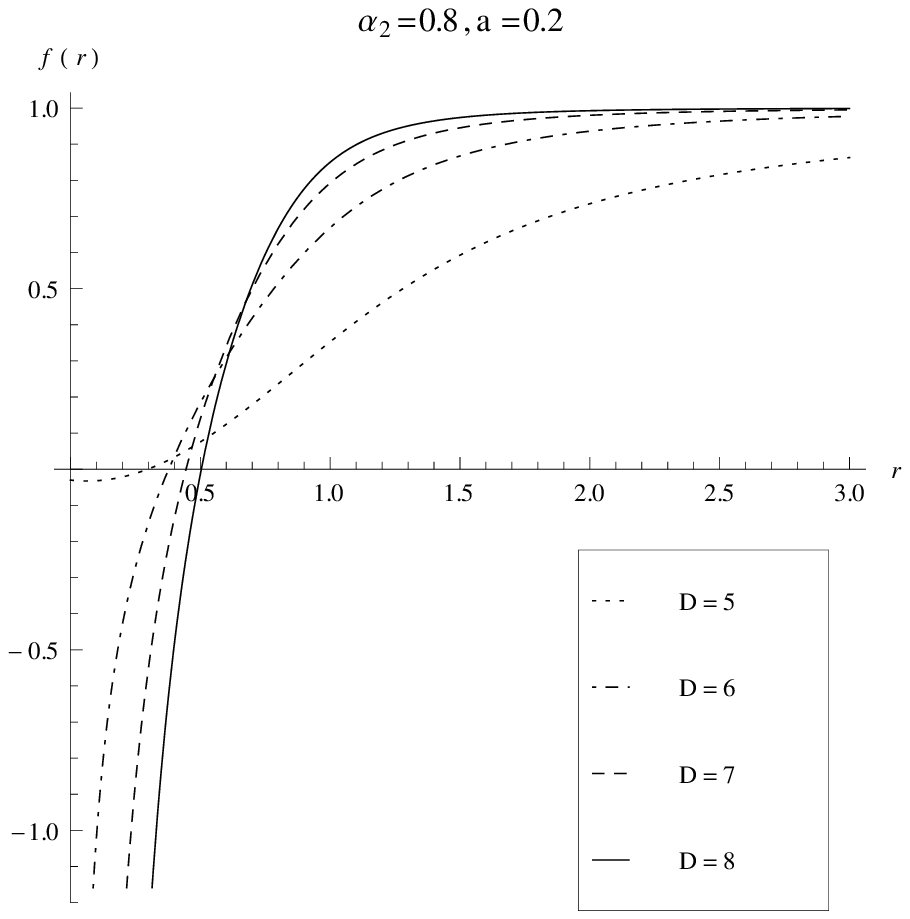}
\\
\hline
\includegraphics[width= 8 cm, height= 6 cm]{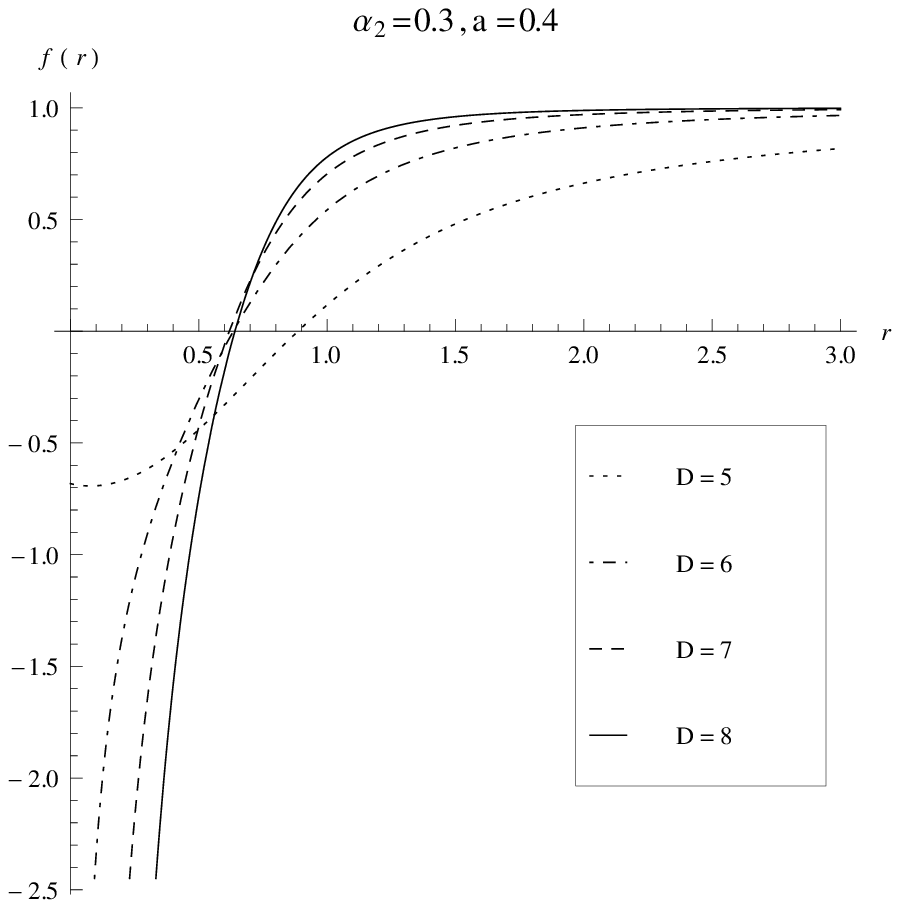}&
\includegraphics[width= 8 cm, height= 6 cm]{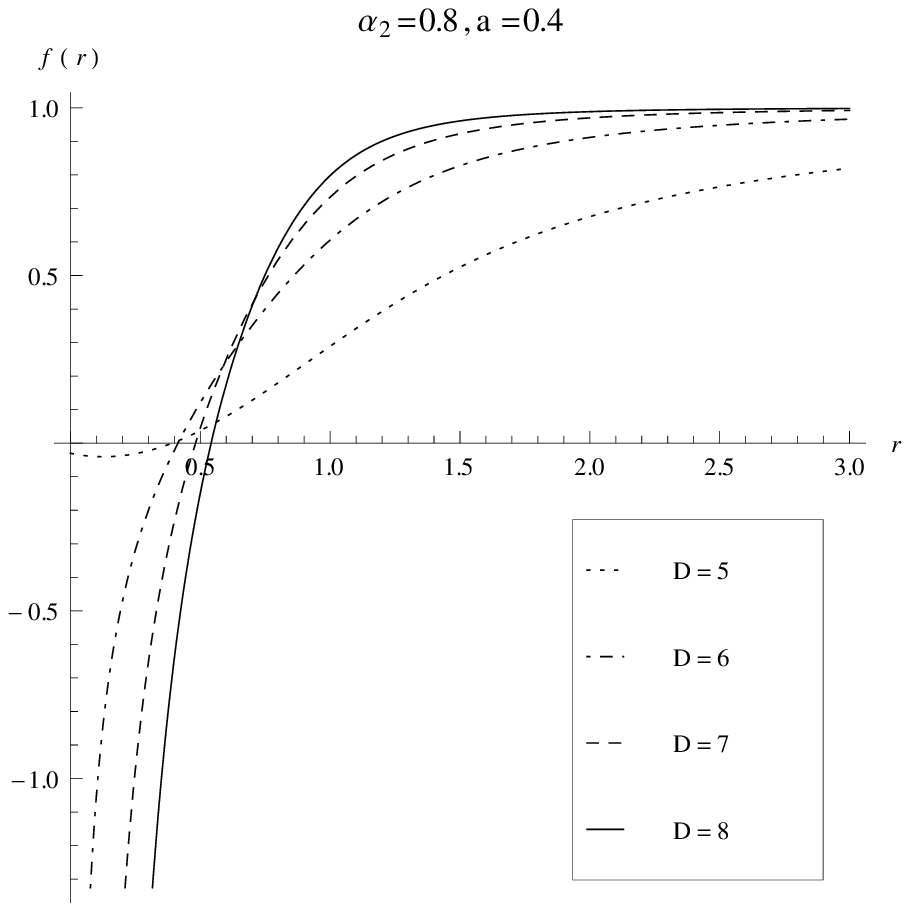}
\\
\hline
\includegraphics[width= 8 cm, height= 6 cm]{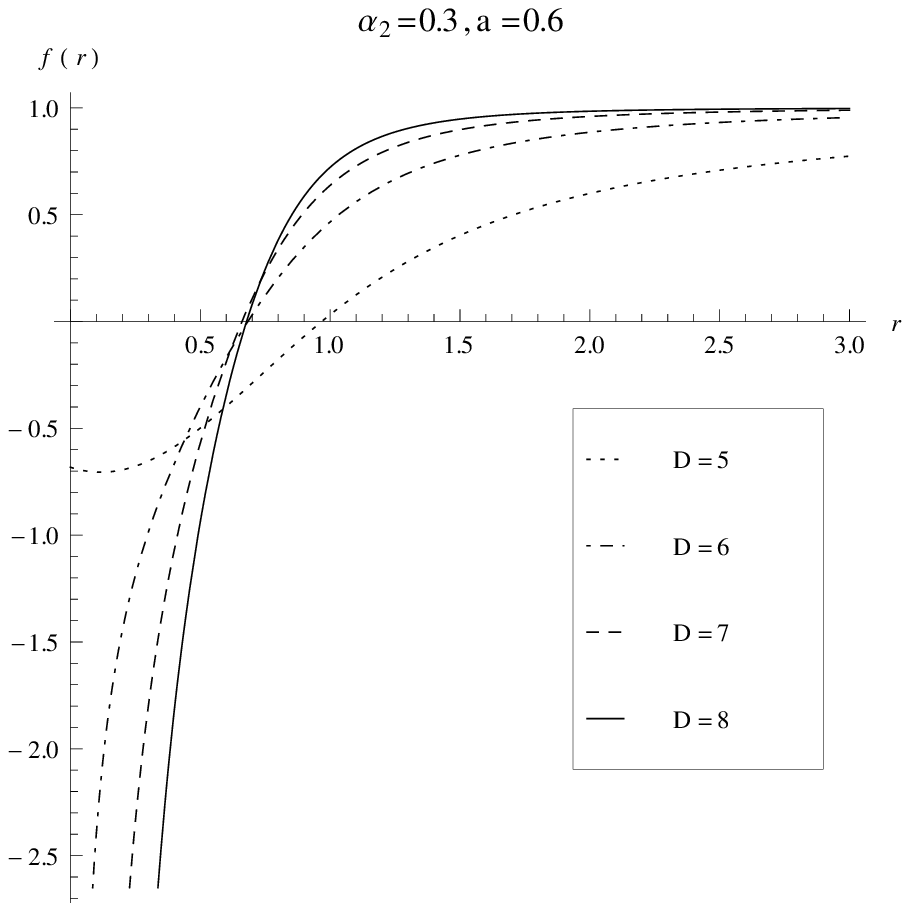}&
\includegraphics[width= 8 cm, height= 6 cm]{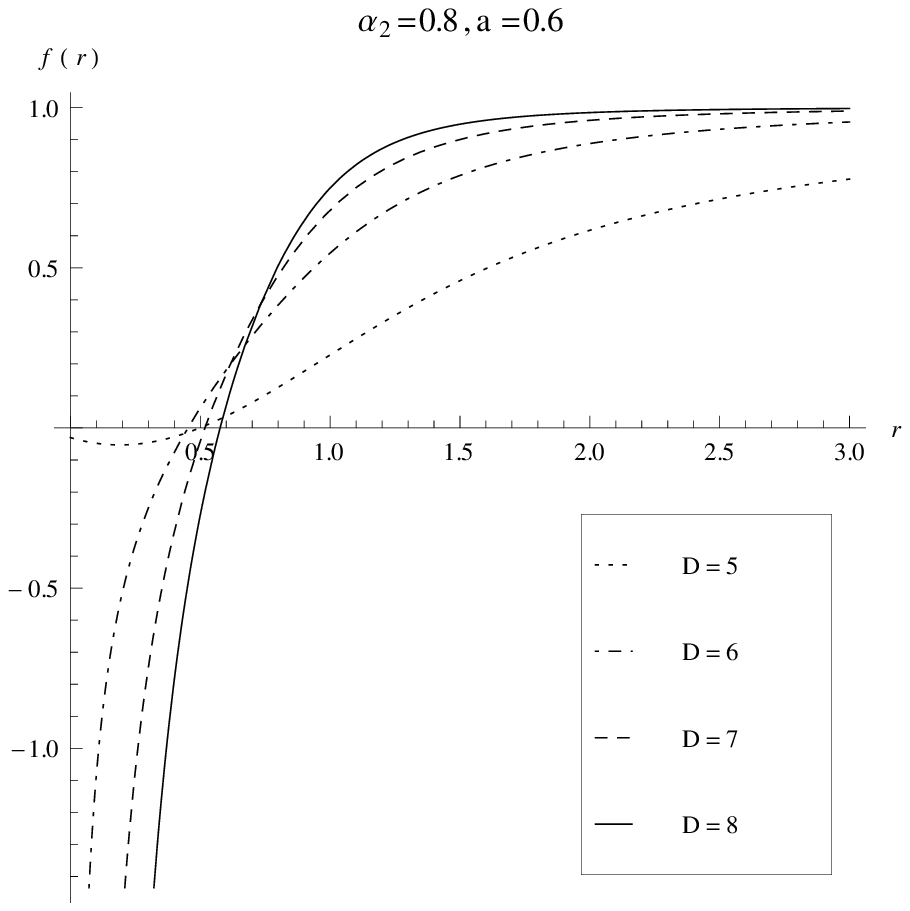}
\\
\hline
\includegraphics[width= 8 cm, height= 6 cm]{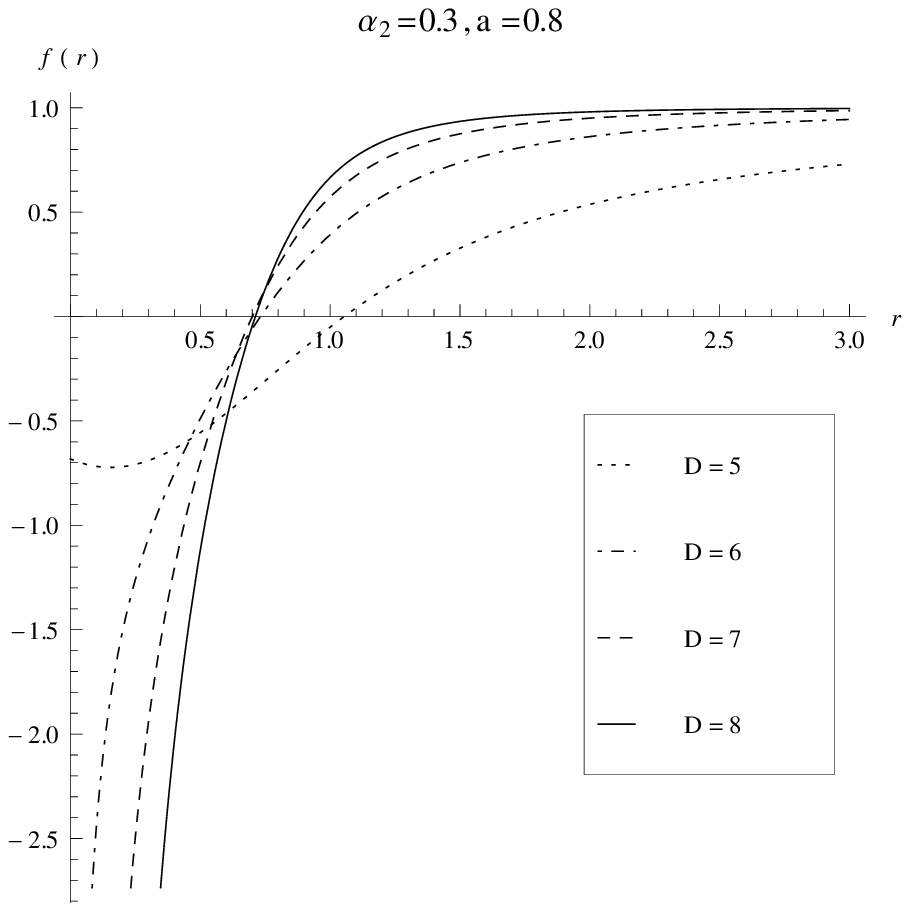}&
\includegraphics[width= 8 cm, height= 6 cm]{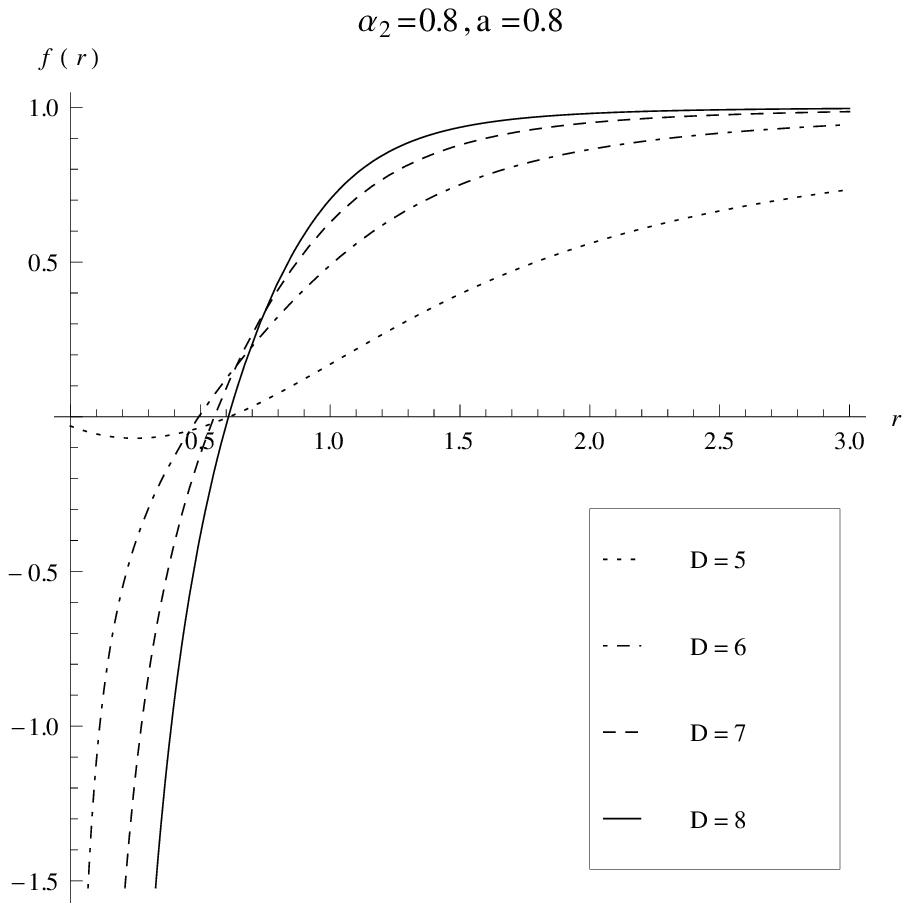}
\\
\hline
\end{tabular}
\caption{\label{f(r) Vs radius} Plot of metric function $f(r)$ vs $r$ in various dimensions with $M=1$ for the Einstein-Gauss-Bonnet case. (Left): $\alpha_2=0.3$ and $a=0.2, 0.4, 0.6$, and $0.8$ (top to bottom). (Right): $\alpha_2=0.8$ and $a=0.2, 0.4, 0.6$, and $0.8$ (top to bottom).}
\end{figure*}
The Hawking temperature associated with the black hole is defined by $T=\kappa/2\pi$, where $\kappa$ is the surface gravity defined by
\begin{equation}
\kappa^{2}=-\frac{1}{4}g^{tt}g^{ij}g_{tt,i}\;g_{tt,j},
\end{equation}
which on inserting the metric function becomes
\begin{equation}
\kappa=\left\vert \frac{1}{2}f^{\prime}(r_+)  \right\vert.
\end{equation}
Accordingly, the Hawking temperature of the black hole on the outer horizon reads
\begin{eqnarray}
T = \frac{\kappa}{2\pi}=\frac{(n-1)}{4 \pi r_+} \left[1-\frac{2a}{n(n-1)r_{+}^{n-2}}\right]. \label{temp}
\end{eqnarray}Then, we can easily see that the temperature is positive for the case $2a<n(n-1)r_{+}^{n-2}$ and negative otherwise. The temperature goes over to zero when $2a=n(n-1)r_{+}^{n-2}$. Taking the limit $a\rightarrow0$, we recover the temperature for higher-dimensional general relativity: \begin{equation}
T= \frac{(n-1)}{4\pi r_+}. \label{tem}
\end{equation}Another useful and important thermodynamic quantity associated with the black hole horizon is its entropy. The black hole behaves as a thermodynamic system; quantities associated with it must obey the first law of thermodynamics $dM=TdS$. Hence, the entropy is given by
\begin{eqnarray}
S = \int {T^{-1} dM} = \int {T^{-1}\frac{\partial M}{\partial r_+} dr_+}, \label{entr}\end{eqnarray}
and substituting (\ref{mass}) and (\ref{temp}) into (\ref{entr}), we arrive at  \begin{eqnarray} S = \frac{(n-1) V_n }{8}r_{+}^{n}. \label{ent}
\end{eqnarray}
We note that $V_n r_{+}^{n}=V_{D-2}r_{+}^{D-2}$ is just the horizon area of a black hole. We therefore conclude that the higher-dimensional black hole also obeys an area law. In the limit $D\rightarrow4$, it becomes the standard area law. It is interesting to note that the formula (\ref{ent}) is independent of a string cloud background.

Next, we turn our attention to the stability of the black holes by computing the specific heat and to study the effect of a string cloud background on the stability of the black hole. It is well known that the thermodynamic stability of the system is related to the sign of the heat capacity.  If the heat capacity is positive, then  the black hole is stable; when it is negative, the black hole is said to be unstable. The heat capacity of the black hole is defined as   \begin{equation}
C = \frac{\partial{M}}{\partial{T}}= \left(\frac{\partial{M}}{\partial{r_+}}\right)\left(\frac{\partial{r_+}}{\partial{T}}\right). \label{SH}
\end{equation} Using Eqs. (\ref{mass}) and (\ref{temp}) in (\ref{SH}), we get
\begin{eqnarray}
C =  -\frac{n  V_n }{8}r_{+}^{n}\left[\frac{n(n-1) r_{+}^{n-2} -2 a}{n r_{+}^{n-2}-2a}\right].\label{SH2}
\end{eqnarray}It is clear that the heat capacity $C$ of the black hole depends on a string cloud parameter $a$. In the limit $a\rightarrow0$, we obtained specific heat of a Schwarzschild-Tangherlini black hole,  \begin{eqnarray}
C =  - \frac{n(n-1) V_n }{8}r_{+}^n. \label{SH3}
\end{eqnarray}
which indicates the thermodynamic instability of the black holes. The extension of the above analysis of Lovelock gravity is an interesting subject to explore.
\section{Einstein-Gauss-Bonnet solutions}
Next, let us consider the theory with $\alpha_3 =0 $, which is usually referred to as the Einstein-Gauss-Bonnet gravity. The static spherically symmetric black hole solution of Einstein-Gauss-Bonnet theory was first obtained by Boulware and Deser \cite{bd}. The simplest Lovelock Lagrangian contains the well-known Gauss-Bonnet term that embodies  nontrivial  dynamics for the gravitational field in five-(and higher-)dimensional theories. Equation~(\ref{master}) with $\alpha_{3}=0$ takes the form%
\begin{eqnarray}
& & n\Big[\left( r^{3}-2\tilde{\alpha}_{2}r\left(  f\left(  r\right)  -1\right)
\right)  f^{\prime}\left(  r\right)  +  \left(  n-1\right)  r^{2}\left(  f\left(  r\right)  -1\right) \nonumber \\
& &  -\left(
n-3\right)  \tilde{\alpha}_{2}\left(  f\left(  r\right)  -1\right)
^{2}\Big]  =\frac{2 a}{r^{n-4}},\nonumber
\end{eqnarray}
which may be called the Einstein-Gauss-Bonnet master equation. This
equation admits a general solution in  arbitrary dimensions as follows%
\begin{equation} \label{sol:egb}
f_{\pm}\left(  r\right)  =1+\frac{r^{2}}{2\tilde{\alpha}_{2}}\left(  1\pm
\sqrt{1+\frac{8\tilde{\alpha}_{2}m}{(n-1)r^{n+1}}+\frac{8\tilde{\alpha}_{2} a}{n  r^{n}}}\right)  ,\text{ \ }n>3.
\end{equation}
The sign $\pm$ refers to the two  different branches of solutions. But only negative (-ve) branch is connected to standard Einstein-Hilbert gravity, as it reduces to the general relativity solution (\ref{ndsol}) when $\alpha_2 \rightarrow 0$.  The above solution is analyzed for the five-dimensional case in Refs. \cite{hr,sgr}. To study the general structure of solution (\ref{sol:egb}), we take the limit $r\rightarrow\infty$ or $m=a=0$ in solution (\ref{sol:egb}) to obtain
\begin{equation}
\lim_{r\rightarrow\infty} f_+(r)= 1+\frac{r^2}{\alpha_2},\;\;\; \lim_{r\rightarrow\infty} f_-(r)= 1;\end{equation}
this means the plus (+) branch of the solution (\ref{sol:egb}) is asymptotically de Sitter (anti-de Sitter) depending on the sign of $\alpha_2$ $(\pm)$, whereas the minus branch of the solution (\ref{sol:egb}) is asymptotically flat. In the large $r$ limit, Eq.~(\ref{sol:egb}) reduces to solution (\ref{ndsol}), and the metric becomes $D$-dimensional Schwarzschild in a string cloud background. In Fig.~\ref{f(r) Vs radius}, we plot the $f(r)$ as a function of $r$ in the various dimensions. It is interesting to note that these solutions admit only one horizon and the radius of the horizon is increasing with the increase in the value of a string cloud parameter $a$. Henceforth, we shall restrict ourselves to the negative branch of the solution (\ref{sol:egb}).  It may be noted that in Eq.~(\ref{sol:egb}) $m$ is related to ADM mass $M$ via (\ref{a2}).
\begin{figure*}
\begin{tabular}{|c|c|}
\hline
\includegraphics[width= 8 cm, height= 6 cm]{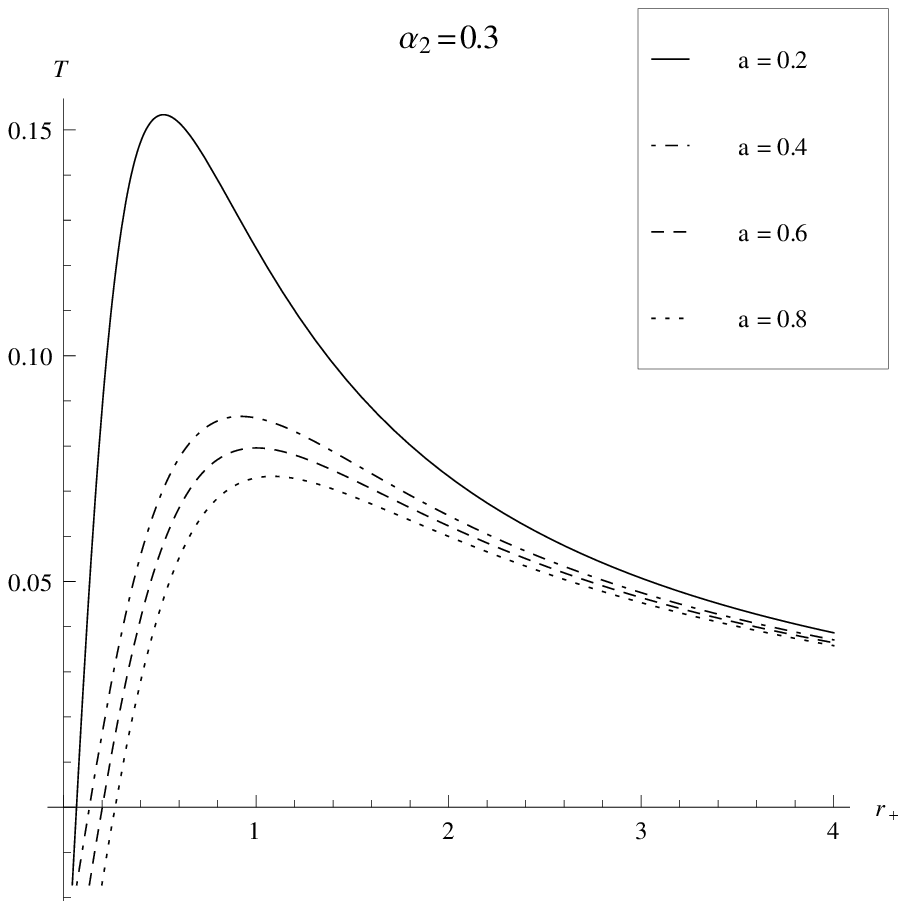}&
\includegraphics[width= 8 cm, height= 6 cm]{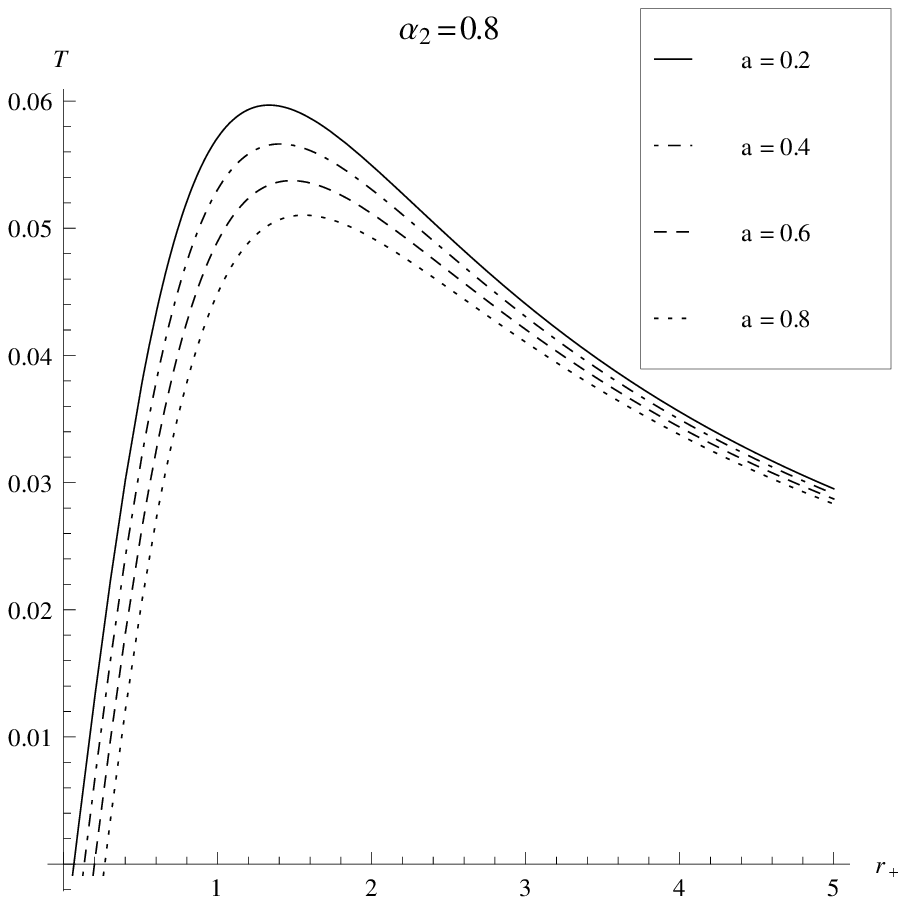}
\\
\hline
\includegraphics[width= 8 cm, height= 6 cm]{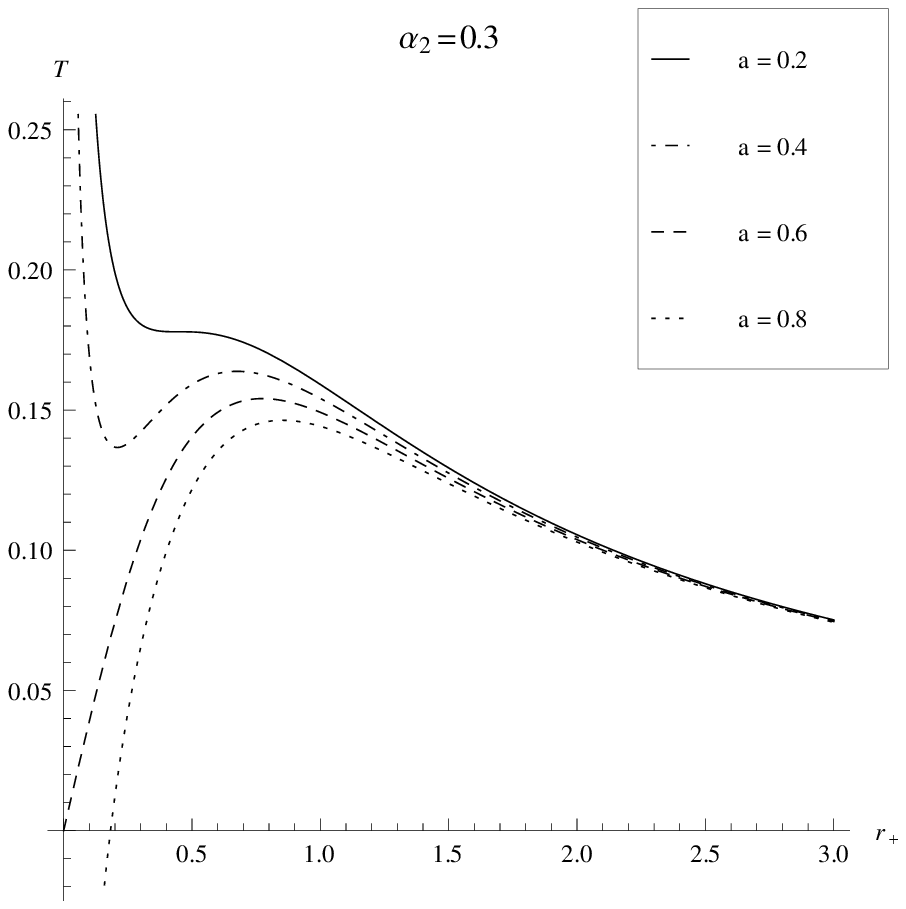}&
\includegraphics[width= 8 cm, height= 6 cm]{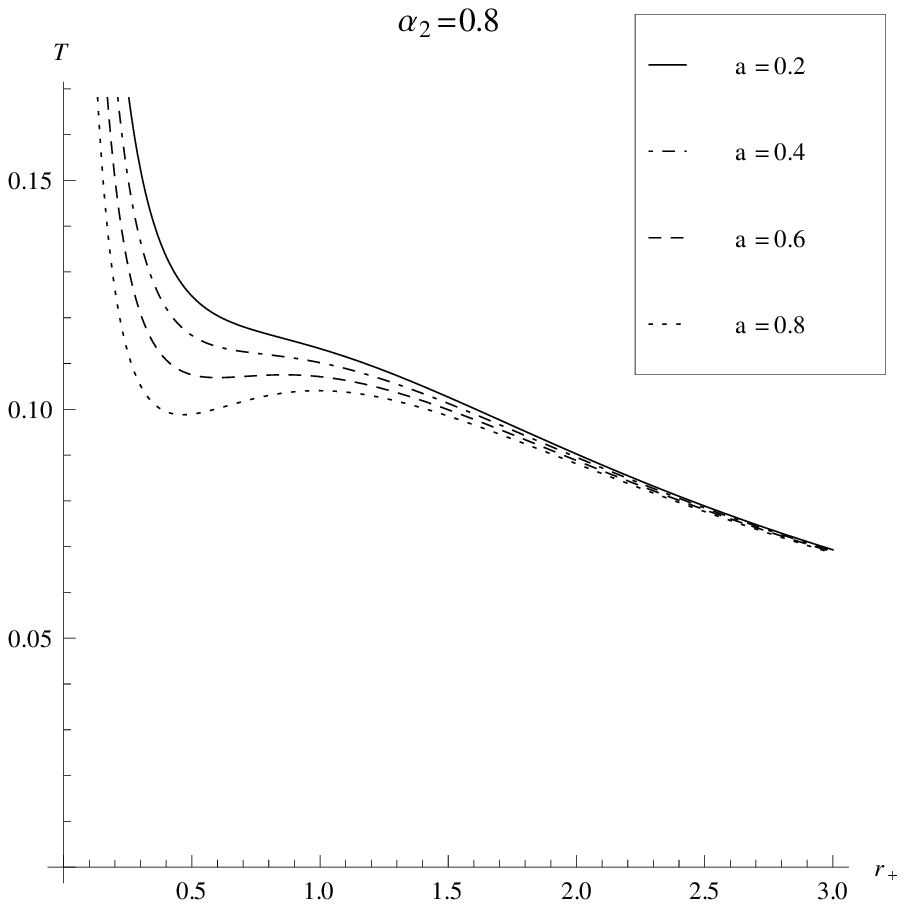}
\\
\hline
\includegraphics[width= 8 cm, height= 6 cm]{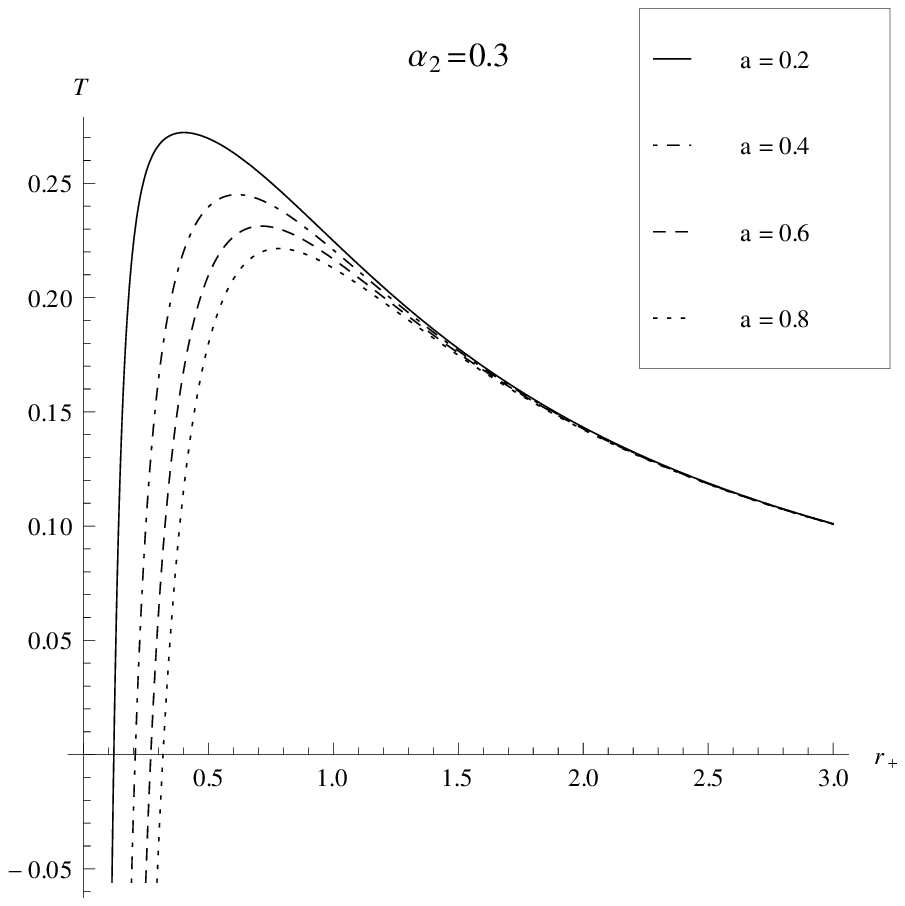}&
\includegraphics[width= 8 cm, height= 6 cm]{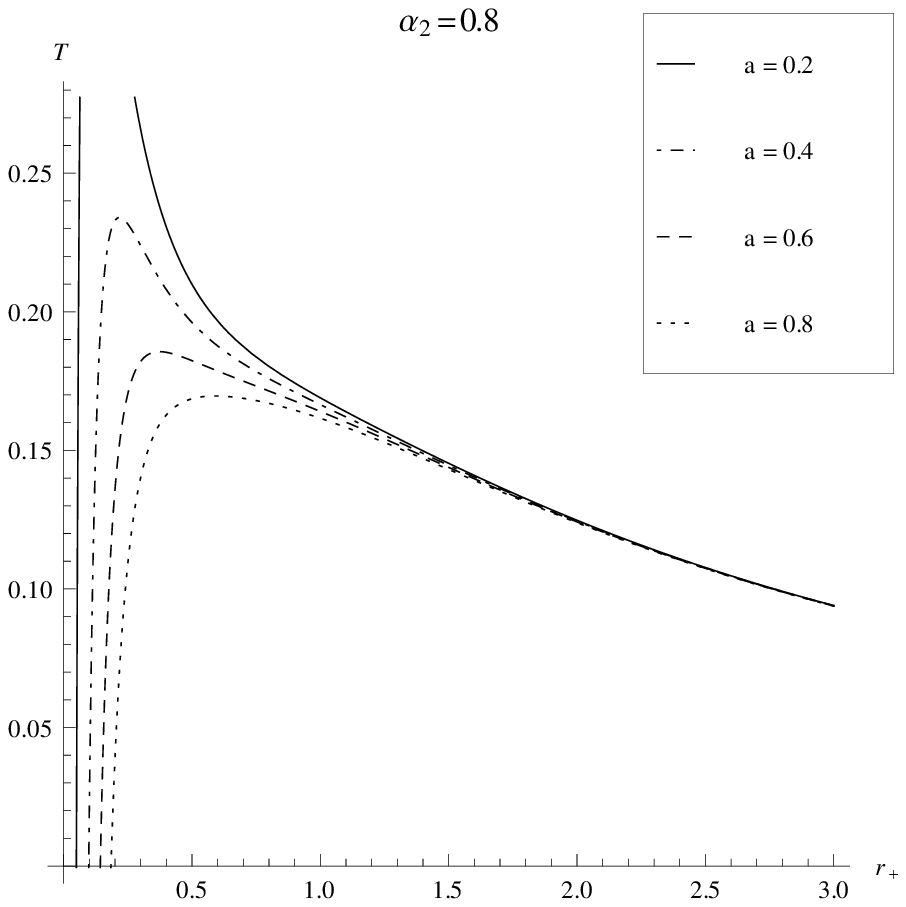}
\\
\hline
\includegraphics[width= 8 cm, height= 6 cm]{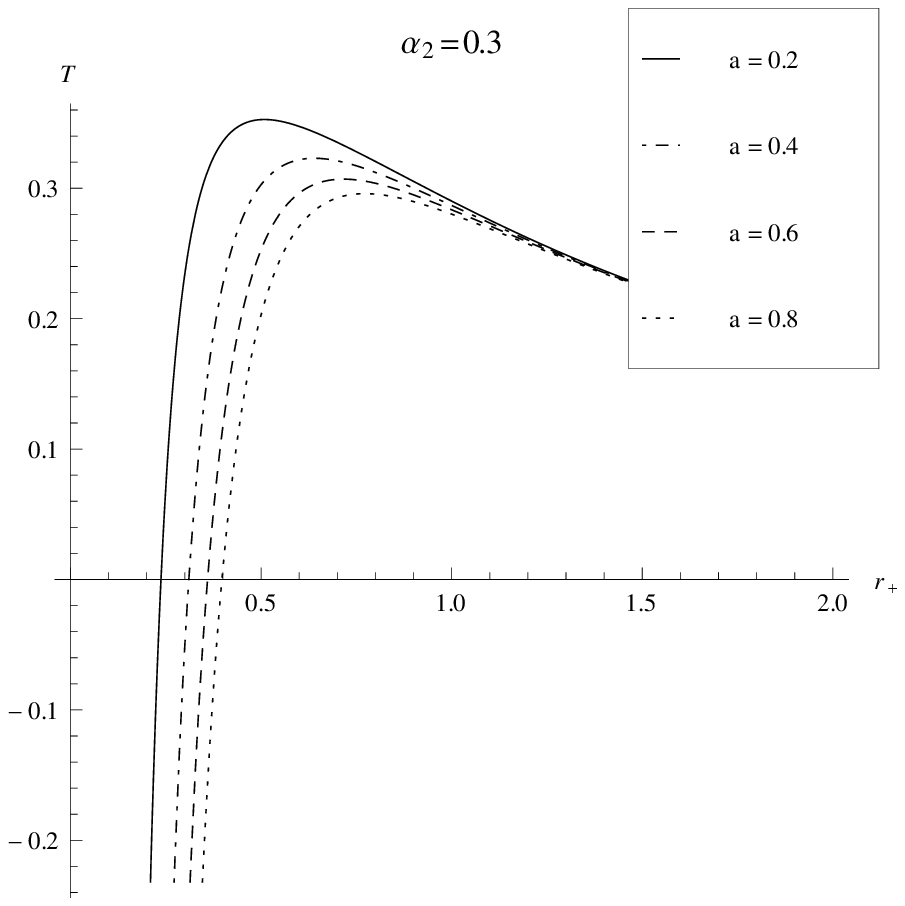}&
\includegraphics[width= 8 cm, height= 6 cm]{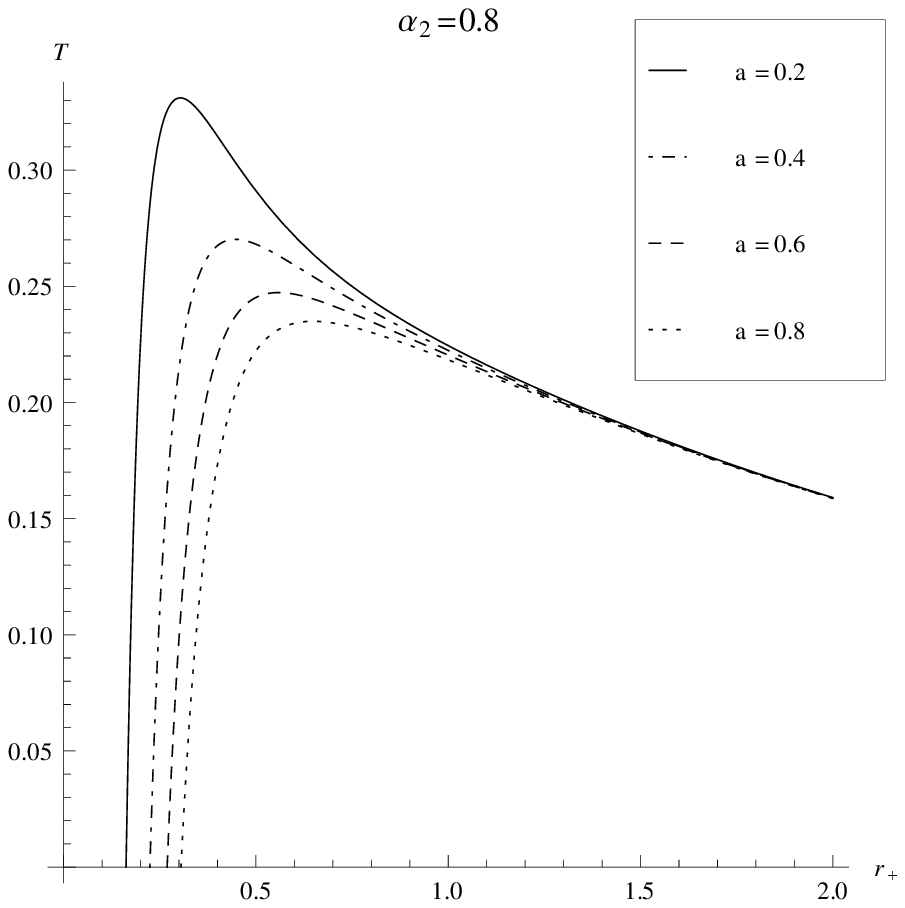}
\\
\hline
\end{tabular}
\caption{\label{TVsR} The temperature vs horizon radius $r_+$ for the Einstein-Gauss-Bonnet case, with $M=1$, in various dimensions $D=5, 6, 7,$ and $8$ (top to bottom). (Left): $\alpha_2=0.3$ and $a=0.2, 0.4, 0.6$, and $0.8$. (Right): $\alpha_2=0.8$ and $a=0.2, 0.4, 0.6$, and $0.8$.}
\end{figure*}

\begin{figure*}
\begin{tabular}{|c|c|}
\hline
\includegraphics[width= 8 cm, height= 6 cm]{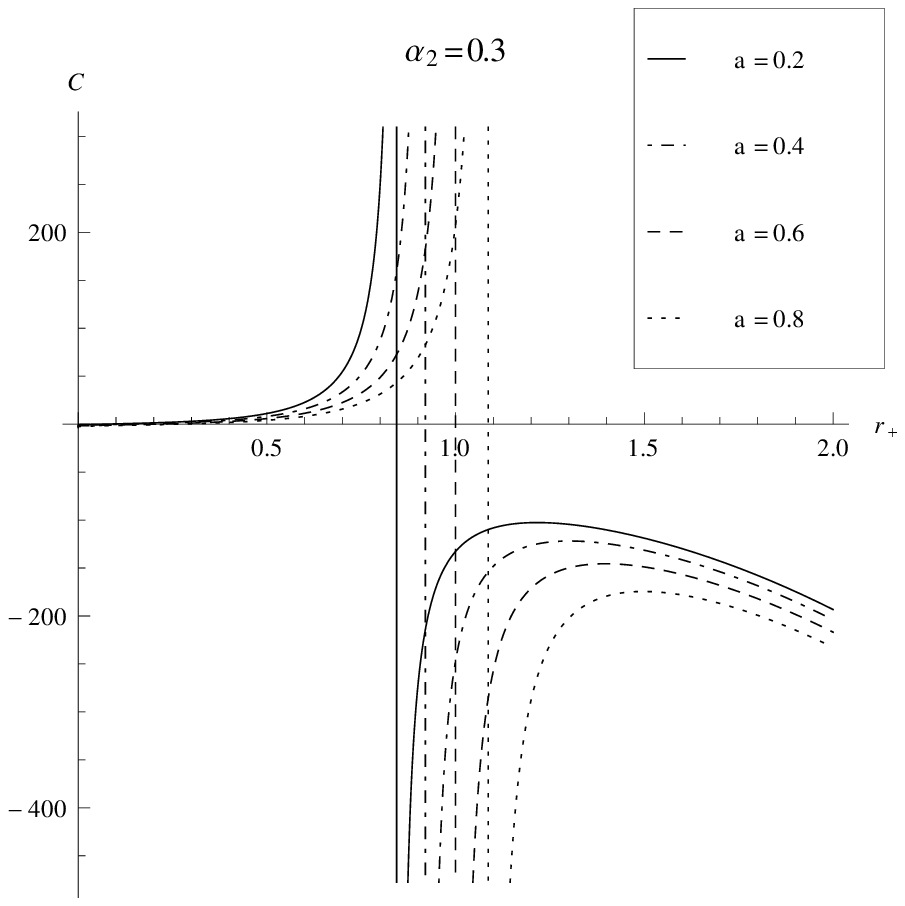}&
\includegraphics[width= 8 cm, height= 6 cm]{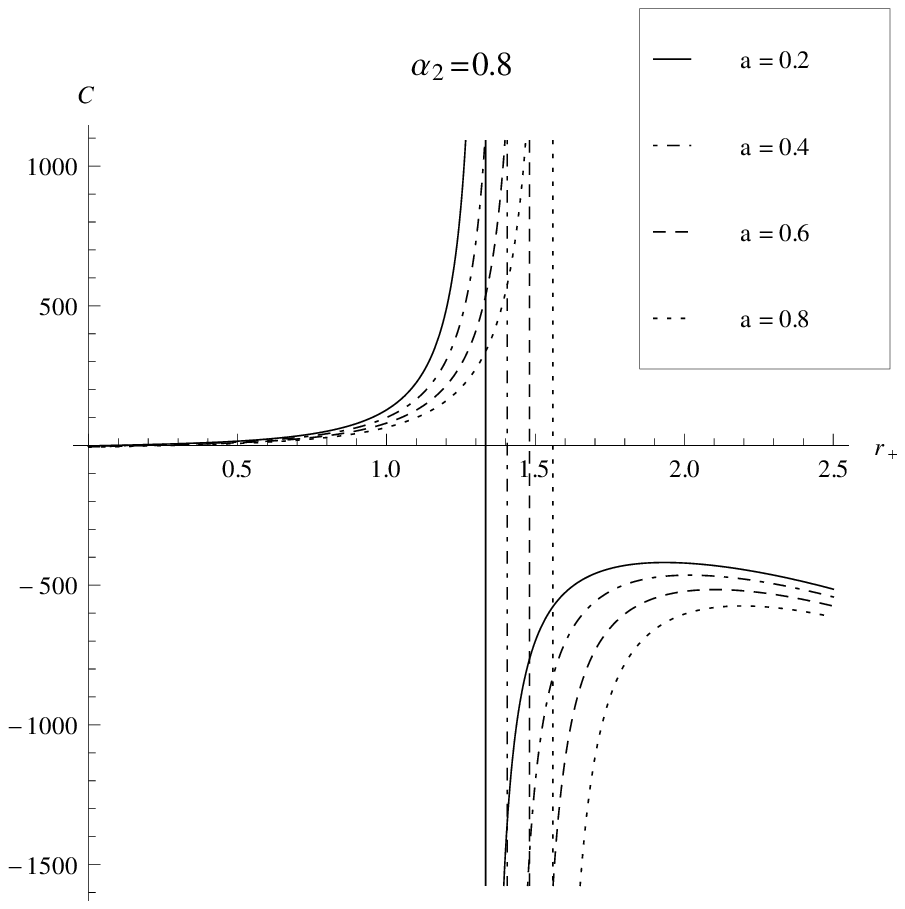}
\\
\hline
\includegraphics[width= 8 cm, height= 6 cm]{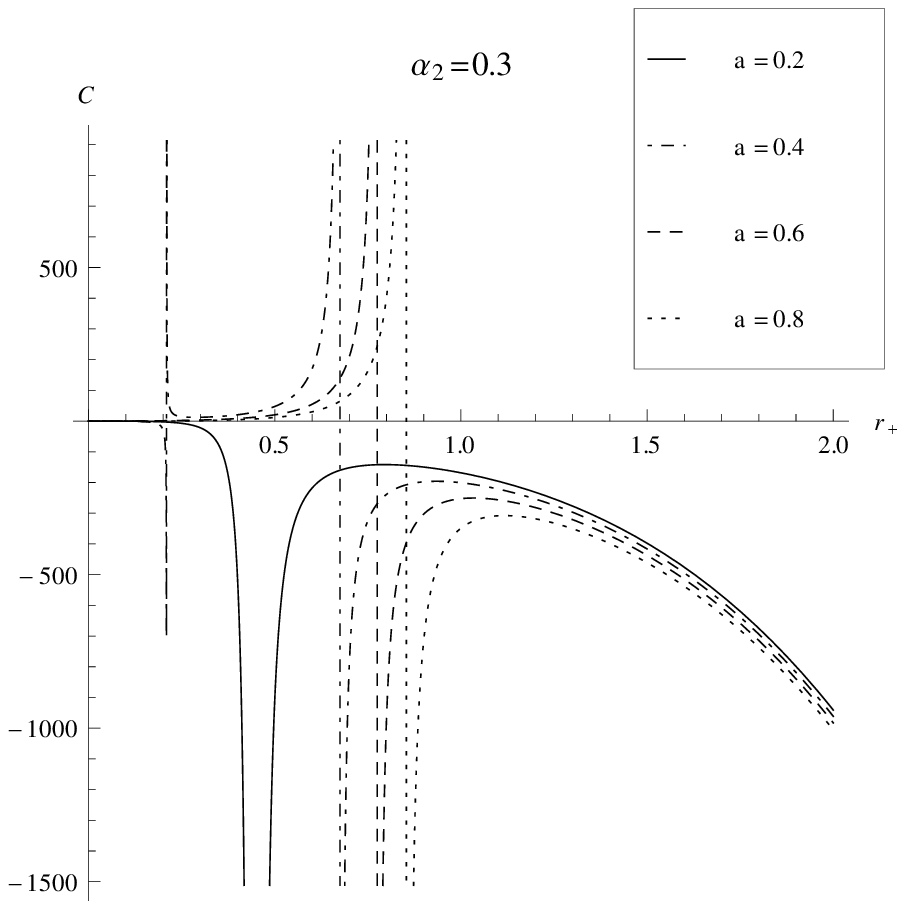}&
\includegraphics[width= 8 cm, height= 6 cm]{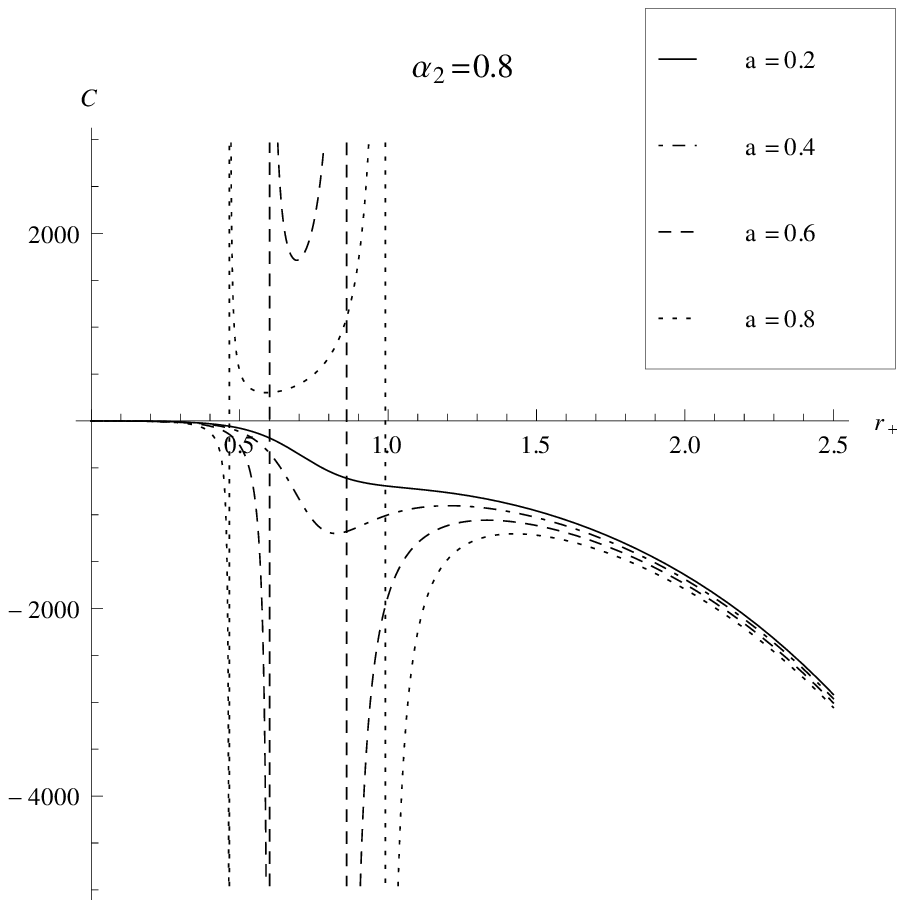}
\\
\hline
\includegraphics[width= 8 cm, height= 6 cm]{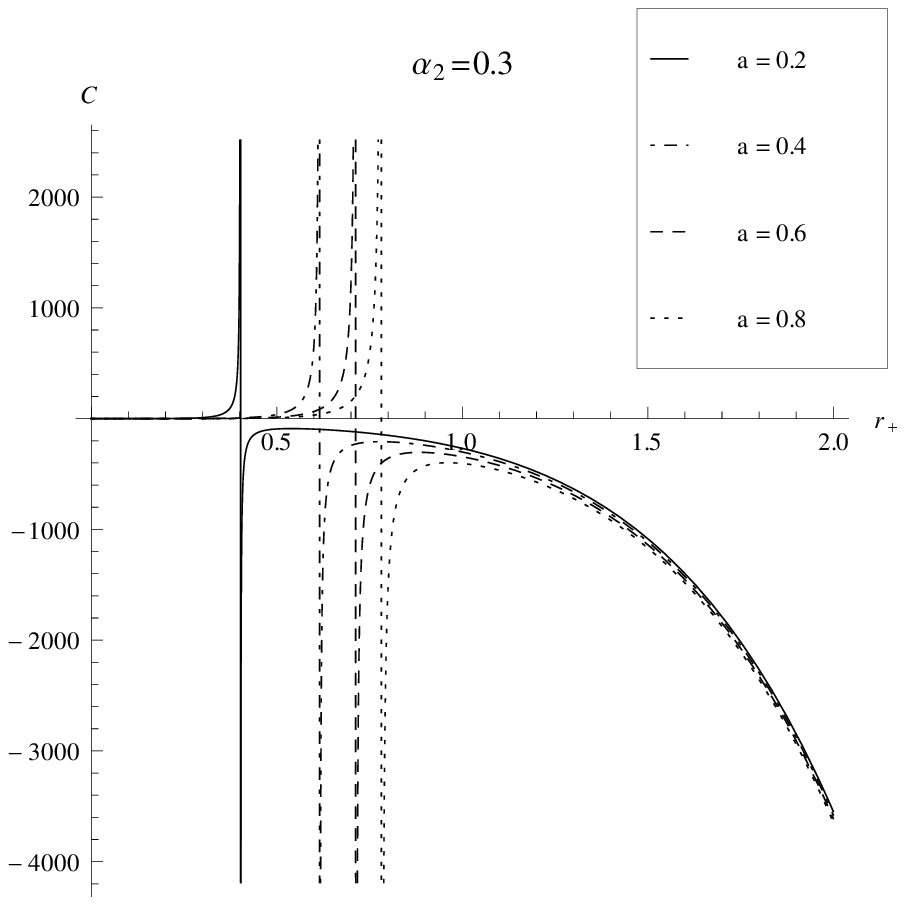}&
\includegraphics[width= 8 cm, height= 6 cm]{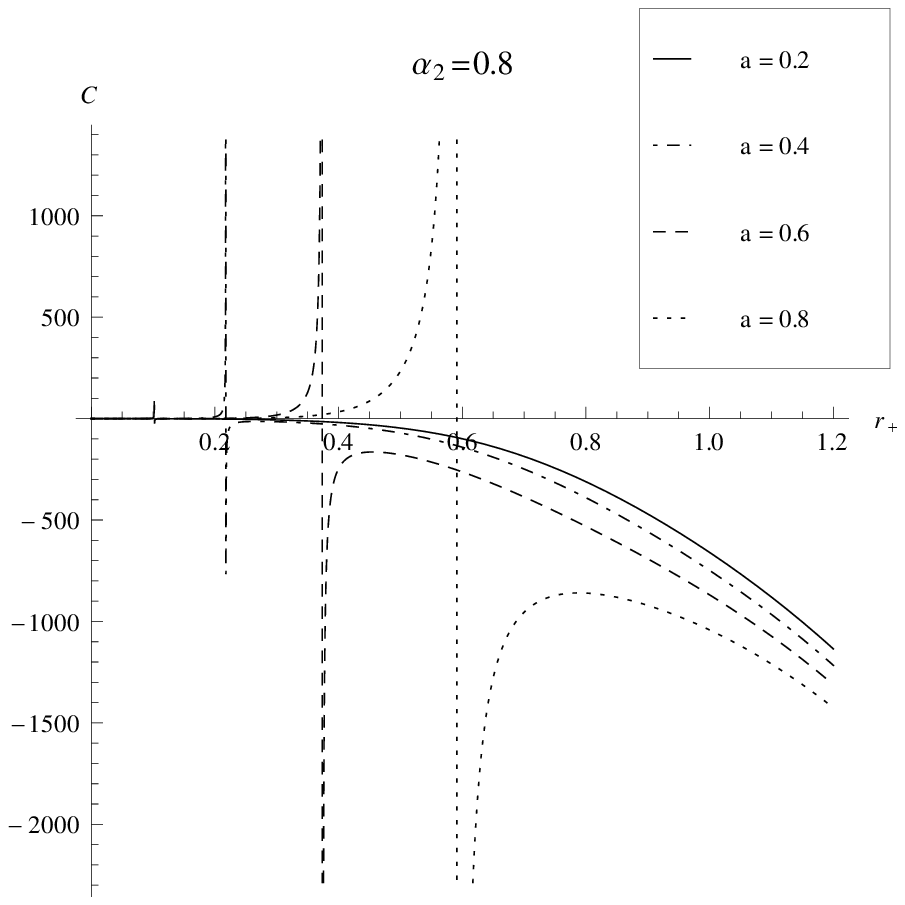}
\\
\hline
\includegraphics[width= 8 cm, height= 6 cm]{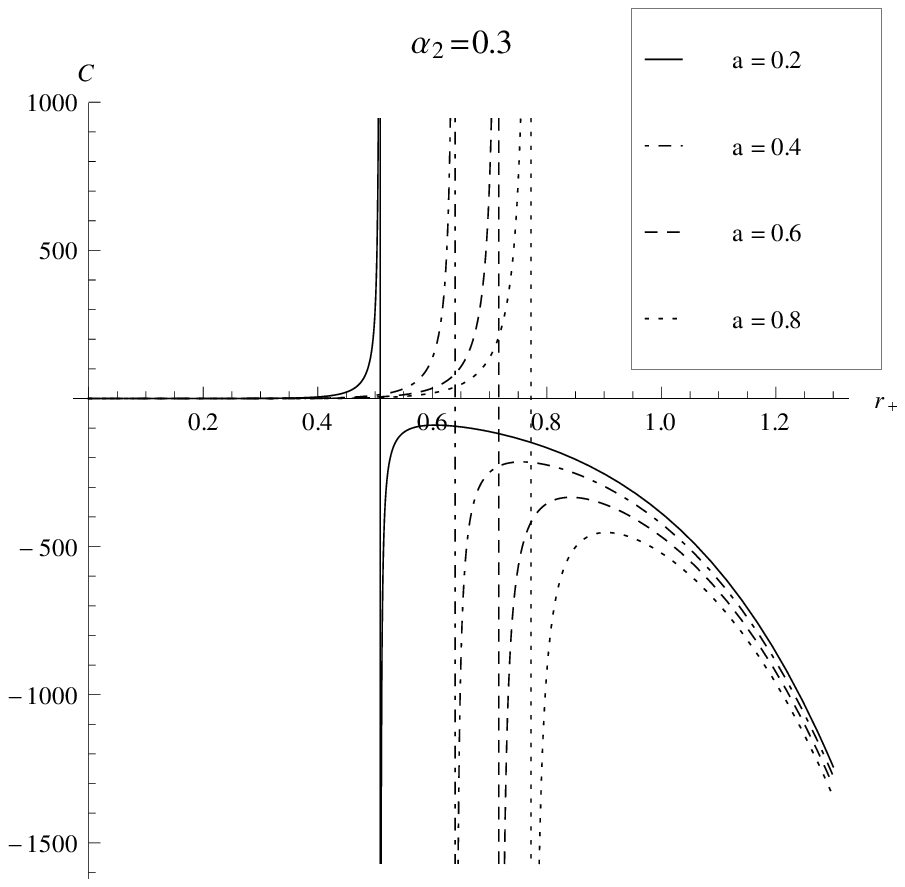}&
\includegraphics[width= 8 cm, height= 6 cm]{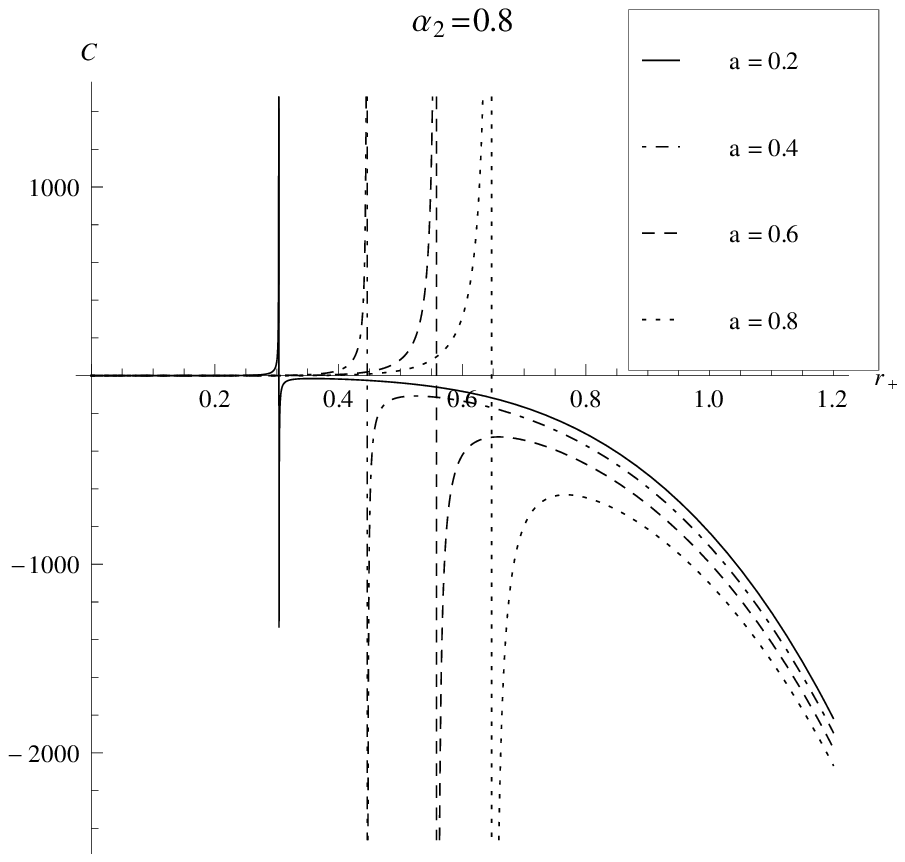}
\\
\hline
\end{tabular}
\caption{\label{CVsR}The specific heat vs horizon radius $r_+$ for the Einstein-Gauss-Bonnet case, with $M=1$, in various dimensions $D=5, 6, 7,$ and $8$ (top to bottom). (Left): $\alpha_2=0.3$ and $a=0.2, 0.4, 0.6$, and $0.8$. (Right): $\alpha_2=0.8$ and $a=0.2, 0.4, 0.6$, and $0.8$.}
\end{figure*}
\section{Thermodynamics of Einstein-Gauss-Bonnet black holes} In this section, we explore the thermodynamics of the Einstein-Gauss-Bonnet black hole solutions (\ref{sol:egb}). The Einstein-Gauss-Bonnet black holes in a string cloud background are characterized by their mass $(M)$ and a string cloud parameter $(a)$.
The mass of the black hole is determined by using $f(r_+)=0$:
\begin{eqnarray}
M_{EGBS} = \frac{n(n-1)  V_n }{32 \pi} r_{+}^{n-1} \left[1 + \frac{\alpha_2}{r_{+}^2} -\frac{2 a}{ n r_{+}^{n-2}}\right]. \label{M1}
\end{eqnarray}In the absence of a string cloud background $(a\rightarrow0)$, we recover the mass obtained for the Gauss-Bonnet black hole: \begin{equation}
M_{EGB} = \frac{n(n-1) V_n }{32 \pi}  r_{+}^{n-1} \left[1 + \frac{\alpha_2}{r_{+}^2}\right].
\end{equation}
To calculate the thermodynamic quantities for the metric (\ref{metric}) with function (\ref{sol:egb}), we use the same approach that was applied in the previous section for the general relativity case. The temperature for the Einstein-Gauss-Bonnet black hole in a string cloud background can be put in the form
 \begin{eqnarray}
T_{EGBS} &=& \frac{(n-1)}{4 \pi r_+} \left[\frac{r_{+}^2+\frac{(n-3)}{(n-1)}\alpha_2-\frac{2a}{n(n-1)r_{+}^{n-4}}}{ (r_{+}^2+2\alpha_2)}\right]. \label{temp1}
\end{eqnarray}
Note that the last factor in Eq.~(\ref{temp1}) modifies the Gauss-Bonnet black hole temperature \cite{Wij}, and taking the limit ${a\rightarrow0}$, we recover it. The Gauss-Bonnet black hole temperature in the absence of a string cloud reads
\begin{equation}
T_{EGB}=\frac{n-1}{4\pi r_+}\left[\frac{r_{+}^2+\frac{(n-3)}{(n-1)}\alpha_2}{r_{+}^2+2\alpha_2}\right],
\end{equation} and when $\alpha_2\rightarrow0$, it becomes the temperature given by Eq.~(\ref{tem}). In Fig.~\ref{TVsR}, we have plotted temperature as a function of $r_+$ in various dimensions. It is interesting to note that for a particular radius of horizon Hawking temperature vanishes. As seen from Fig.~\ref{TVsR}, the Hawking temperature exhibits a peak that decreases and moves to the right when a string cloud parameter $a$ grows.

The entropy of a black hole typically satisfies the area law that states that the entropy of a black hole is a quarter of the event horizon area \cite{jdb}. Now, the entropy of the Einstein-Gauss-Bonnet gravity black holes in a string cloud background, determined using Eq.~(\ref{entr}), reads
\begin{eqnarray}
S_{EGBS}= \frac{(n-1) V_n}{8} r_{+}^n\left[1+\frac{n}{n-2}\frac{2\alpha_2 }{r_{+}^2}\right]. \nonumber
\end{eqnarray}The entropy for our model differs from the expression for entropy in general relativity, in which it is proportional to the area of the event horizon. However, it is interesting to note that the entropy of the  black hole has no effect of a background string cloud.

Finally, we analyze how a string cloud background influences  the thermodynamic stability of the Einstein-Gauss-Bonnet black holes. The thermodynamic stability of a black hole is performed by analyzing the behavior of its heat capacity. The heat capacity of Einstein-Gauss-Bonnet black hole in a string cloud model, using Eqs. (\ref{SH}), (\ref{M1}), and (\ref{temp1}), reads
\begin{eqnarray}
C &=& \frac{n(n-1)  V_n}  {\delta_2}  r_{+}^{n-2} \Big[2 a r_{+}^4 \alpha_2 (r_{+}^2+2 \alpha_2)^2- n r_{+}^n  \nonumber \\ && ((n-1)r_{+}^2+(n-3)\alpha_2)(r_{+}^2+2 \alpha_2)^2 \alpha_2\Big],  \label{SH1}
\end{eqnarray}with
\begin{eqnarray}
\delta_2 &=& 8 \Big[2ar_{+}^4\alpha_2(r_{+}^2+6\alpha_2)+n^2r_{+}^n\alpha_2(r_{+}^4+r_{+}^2\alpha_2+2\alpha_2^2)\nonumber \\ &&-n(r_{+}^{n+4}\alpha_2 +7r_{+}^{n+2}\alpha_2^2 +6r_{+}^n \alpha_2^3 +4ar_{+}^4 \alpha_2^2 +2ar_{+}^6 \nonumber \\ && \alpha_2)\Big]  .
\end{eqnarray}
It is clear that the heat capacity depends on the Gauss-Bonnet coefficient $\alpha_2$, a string cloud parameter $a$, and the dimensions $D$.  When $\alpha_2\rightarrow0$, it returns to the general relativity case. If in addition $a=0$, it becomes Eq.~(\ref{SH3}). In what follows, we analyze the stability of the Einstein-Gauss-Bonnet black hole and bring out the effect of a string cloud background. It is difficult to analyze the heat capacity analytically due to complexity of  Eq.~(\ref{SH1}); hence, we plot it in Fig.~\ref{CVsR} for different values of $a$, $\alpha_2$, and $D$.
Clearly, the positivity of the heat capacity $C$ is sufficient to ensure thermodynamic stability. Figure~\ref{CVsR} shows that heat capacity is discontinuous exactly at one point for a given value of $a$ and $\alpha_2$, which is identified as the critical radius $r_c$. Further, we note that there is a flip of sign in the heat capacity around $r_c$. Thus, the black hole is thermodynamically stable for  $r_+ < r_c$, whereas it is thermodynamically unstable for $r_+>r_c$, and there is a phase transition at $r_+=r_c$ from the stable to unstable phases. Thus, the heat capacity of an Einstein-Gauss-Bonnet black hole, in any dimension for different values of $a$ and $\alpha_2$, is positive for $r_+ < r_c$, while for $r_+>r_c$, it is negative. It is worthwhile to mention that the critical radius $r_c$ changes drastically in a string cloud model, thereby affecting the thermodynamical stability. Indeed, the value of $r_c$ increases with the increase in the string cloud parameter $a$ for a given value of the Gauss-Bonnet coupling constant $\alpha_2$. On the other hand, the $r_c$ decreases with an increase in $\alpha_2$ in all dimensions for $D>6$, and $r_c$ increases with $\alpha_2$ in $D\leq6$. It is notable that the   black hole is thermodynamically stable when the temperature of black hole satisfies $0<T<T_c$, while it is unstable for $T>T_c$ [$(T_c)$ is critical temperature at the critical radius $r_c$].

\section{Conclusion}
Lovelock theories share the property of general relativity  that no derivatives of the curvature tensor, and thus only second derivatives of the metric tensor, arise in the field equations. It follows that Lovelock gravities share a number of additional nice properties with Einstein gravity that are not enjoyed by other more general higher-curvature theories.  Hence, these theories receive  significant attention, especially when finding black hole solutions.
In this paper, we have obtained exact static spherically symmetric black hole solutions to  general relativity, Einstein-Gauss-Bonnet gravity, and the third-order Lovelock gravity (see the Appendix) in the background of a cloud of strings in arbitrary $D=n+2$ dimensions. Thus, we have generalized the static, spherically symmetric black hole solutions for these theories in a string cloud background. We then proceeded to find exact expressions, in the Einstein-Gauss-Bonnet gravity,  for the thermodynamic quantities like the black hole mass, Hawking temperature, and entropy, and in turn also analyzed the thermodynamic stability of black holes for  the case of first- and second-order theories. In addition, we explicitly brought out the effect of a background string cloud on black hole solutions and their thermodynamics.

In particular, these black holes are thermodynamically stable with a positive heat capacity for the range $0 < r < r_c$, and their entropy does not obey a horizon area formula.  Interestingly,  the  Einstein-Gauss-Bonnet black hole entropy has no correction from a string cloud background.  The possibility of a further generalization of these results in arbitrary dimensional Lovelock gravity is an interesting problem for future research.

\acknowledgements
S.G.G. and U.P. thank the University Grant Commission (UGC) for the major research project Grant F. N.  39-459/2010 (SR). S.D.M. acknowledges that this work is based upon research supported by South African Research Chair Initiative of the Department of Science and Technology and the National Research Foundation.

\appendix*
\section{general case,  $\alpha_{3}\neq 0$ and $\alpha_{2} \neq 0 $ in $D$- Dimensions}Here, we study exact solutions of general Lovelock theories for a  string cloud source in arbitrary dimensions and look for particular solutions. Again, it is enough to solve (\ref{master}), which admits a solution,
\begin{figure*}

\begin{tabular}{|c|c|c|c|}
\hline
\includegraphics[width= 8 cm, height= 5 cm]{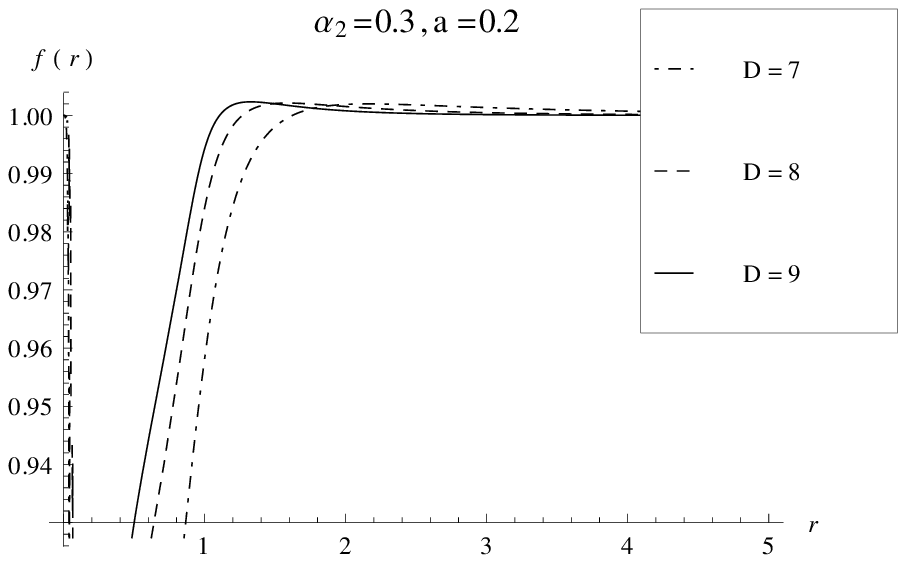}&
\includegraphics[width= 8 cm, height= 5 cm]{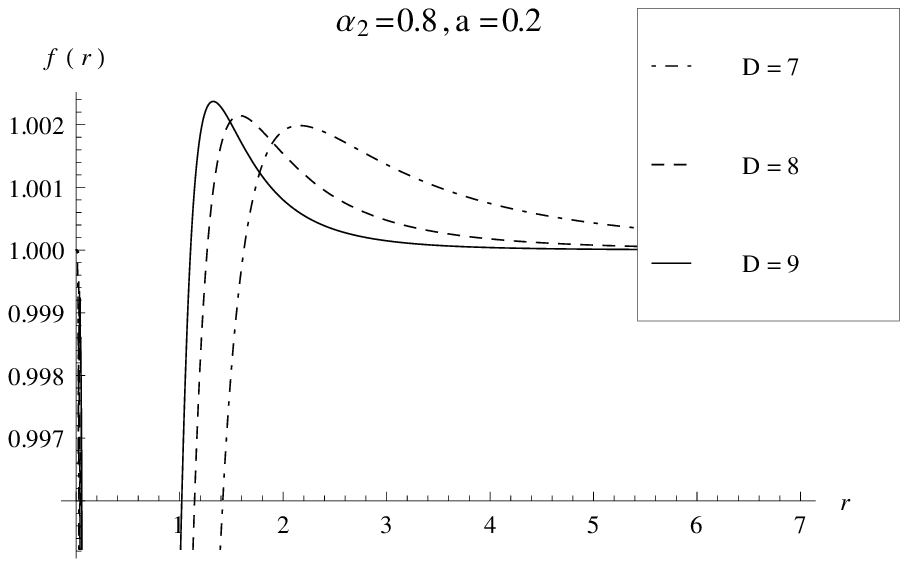}
\\
\hline
\includegraphics[width= 8 cm, height= 5 cm]{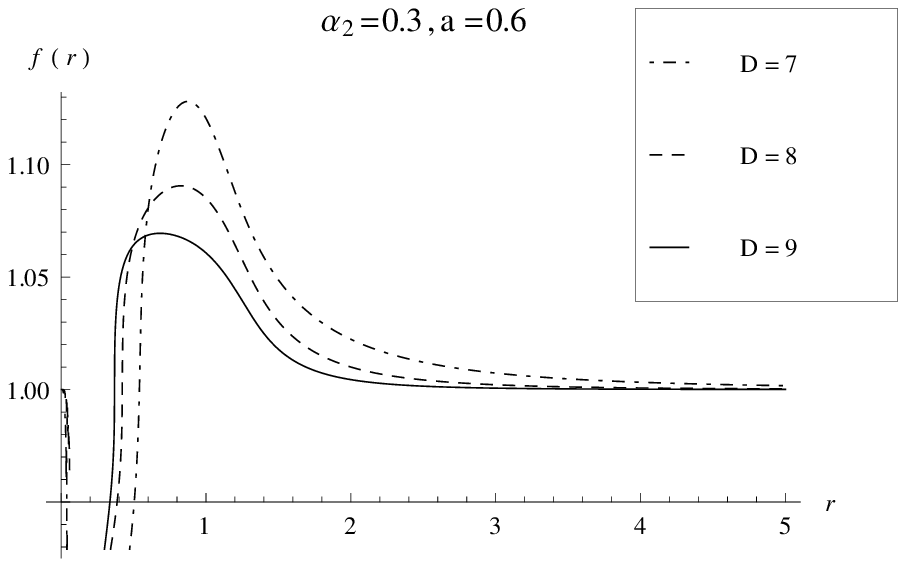}&
\includegraphics[width= 8 cm, height= 5 cm]{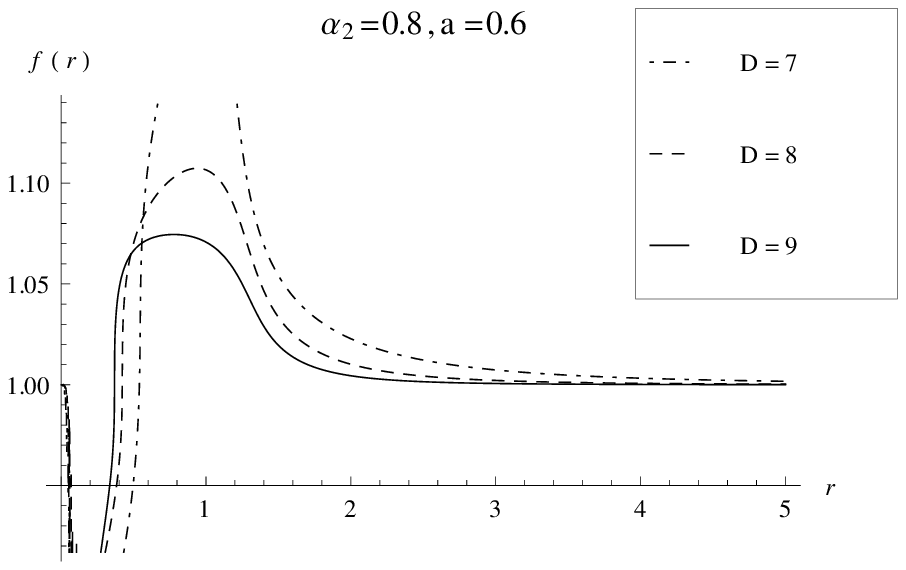}
\\
\hline
\end{tabular}
\caption{\label{EGBLLf(r)}  Plot of metric function $f(r)$ vs $r$ for the general Lovelock gravity ($\alpha_2\neq \alpha_3 \neq 0$) with $\alpha_3=1$ and $M=1$ in various dimensions. (Left): $\alpha_2=0.3$ and $a=0.2$  and $0.6$ (top to bottom). (Right): $\alpha_2=0.8$ and $a=0.2$ and $0.6$ (top to bottom). }
\end{figure*}

 \begin{figure*}

\begin{tabular}{|c|c|c|c|}
\hline
\includegraphics[width= 8 cm, height= 5 cm]{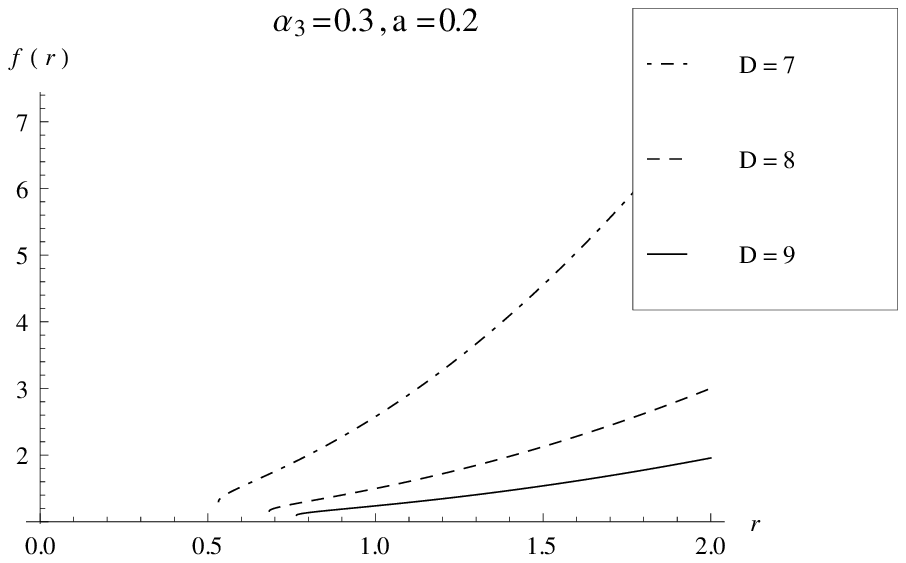}&
\includegraphics[width= 8 cm, height= 5 cm]{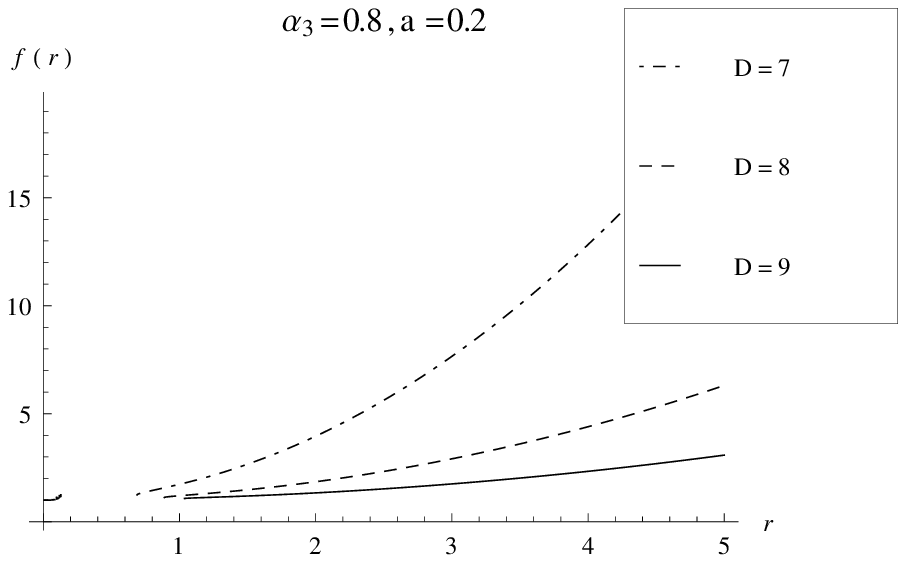}
\\
\hline
\includegraphics[width= 8 cm, height= 5 cm]{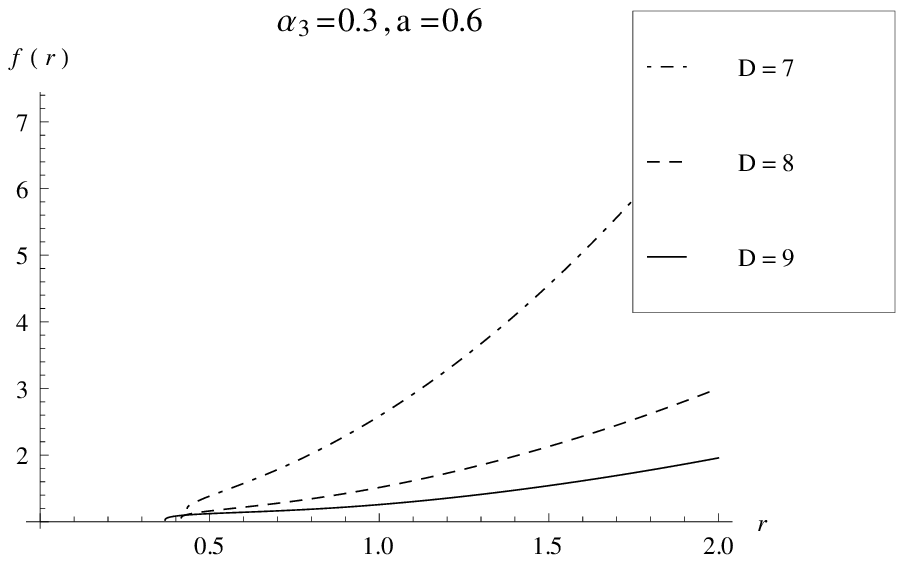}&
\includegraphics[width= 8 cm, height= 5 cm]{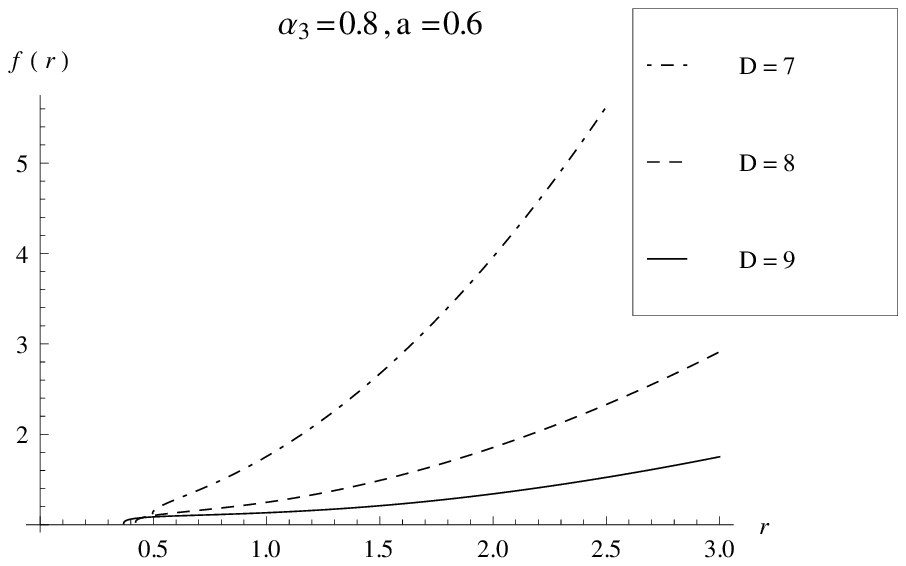}
\\
\hline
\end{tabular}
\caption{\label{EGBLLf(r)1}  Plot of metric function $f(r)$ vs $r$ for the general Lovelock gravity ($\alpha_2\neq \alpha_3 \neq 0$) with $\alpha_2=1$ and $M=1$ in various dimensions. (Left): $\alpha_3=0.3$ and $a=0.2$  and $0.6$ (top to bottom). (Right): $\alpha_3=0.8$ and $a=0.2$ and $0.6$ (top to bottom). }
\end{figure*}

\begin{eqnarray} \label{sol:ll}
f(r)&=&1+ \frac{\tilde{\alpha_2}}{3\tilde{\alpha_3}}r^2 + \epsilon_1(r)
\Delta^{1/3} +\frac{ \epsilon_2(r) }{\Delta^{1/3}}(\tilde{\alpha_2}^2-3\tilde{\alpha_3}),
\end{eqnarray}
where
\begin{eqnarray}
\epsilon_1(r)&=&\frac{1}{3 (2n)^{1/3} \tilde{\alpha}_3 r^{(n-5)/3}},\;  \nonumber \\
\epsilon_2(r) &=&\frac{(2n)^{1/3} r^{(n+7)/3 }}{3 \tilde{\alpha}_3}, \nonumber \\
\Delta  &=&3 \sqrt{3} \sqrt{\delta} \tilde{{\alpha}_3} + 2 \tilde{\alpha_2} \left( {\tilde{{\alpha}_2}}^{2}-\frac{9{\tilde{\alpha}_3} }{2}\right) n r^{(n+1)} \nonumber \\ && + 54 \tilde{\alpha_3}^2 (ar -m), \nonumber \\
\delta &=& -{n}^{2} \left( {{\tilde{\alpha}}_2}^{2}-4\,\tilde{{\alpha}_3} \right) {r}^{2\,n
+2}+8{r}^{n+1}\left(a{r} - m \right) \tilde{{\alpha}_2} \nonumber \\
\times && n \left( {\tilde{{\alpha}}_2}^{2}-\frac{9\tilde{{\alpha}_3} }{2}\right) +108\,{\tilde{{\alpha}_2}}^{2} \left( a\; r - m \right) ^{2},\nonumber
\end{eqnarray}
and $m$ is again an integration constant related to the mass of the black hole.

In Figs.~\ref{EGBLLf(r)} and \ref{EGBLLf(r)1}, we have plotted $f(r)$ for the general Lovelock gravity case $\alpha_2\neq\alpha_3\neq0$ in various dimensions. Figure~\ref{EGBLLf(r)} shows the variation of $f(r)$ with fixed $\alpha_2$ and varying $\alpha_3$; Fig.~\ref{EGBLLf(r)1}, shows the variation of $f(r)$ with fixed $\alpha_3$ and varying $\alpha_2$.

The solution (\ref{sol:ll}) reduces to the Lovelock black hole for $a=0.$ However, the solution is very complex and difficult to analyze, and hence, in what follows, we specialize to the seven-dimensional case.
\paragraph{ $\alpha_{3}\neq 0 \; \alpha_{2} \neq 0 $, \mbox{and} $D=7$:}
It may be noted that in seven dimensions we can see the role of both second-order and third-order Lovelock parameters.
It is interesting to observe that enormous simplification occurs in the above solution in seven dimensions $(D=7)$.  The metric can be written as
\begin{eqnarray}
f(r)&=&1+ \frac{\tilde{\alpha_2}}{3\tilde{\alpha_3}}r^2 + \frac{\Delta^{\frac{1}{3}}}{30 \tilde{\alpha}_3} + \frac{10 r^4(\tilde{\alpha}_2^2 - \tilde{\alpha}_3) }{\Delta^{\frac{1}{3}}},
\end{eqnarray}
where
\begin{eqnarray}
 \Delta &= & 300 \sqrt{3} \tilde{\alpha_3} \sqrt{\delta} + 5400 (ar-m) \tilde{\alpha_3}^2  + 10 \tilde{\alpha_2}^2 r^6 \nonumber \\ && (10 \tilde{\alpha_2}  - 45\tilde{\alpha_3} ), \nonumber \\
 \delta &= & 25 (4\tilde{\alpha_3}-  \tilde{\alpha_2}^2 ) + 40 \tilde{\alpha_2} r^6 \gamma (ar - m) + 108 \tilde{\alpha_3} ^2 \nonumber \\ && (ar-m)^2. \nonumber
\end{eqnarray}
To proceed further, we consider a simple but interesting case of the following.
\paragraph{$\tilde{\alpha_3} =2 \tilde{\alpha_2}^2/9:$} The metric function $f(r)$ reads
\begin{eqnarray}
f(r)&=& 1+ \frac{3r^2}{2\tilde{\alpha_2}} + \frac{3\Delta^{\frac{1}{3}}}{20 \tilde{\alpha_2^2}} + \frac{5 r^4 }{\Delta^{\frac{1}{3}}},
\end{eqnarray}

\begin{eqnarray}
\Delta &=  & \frac{200 \sqrt{16}\tilde{\alpha_2}^2 }{3} \sqrt{\left((ar-m)^2\tilde{\alpha_2}^2  \frac{25r^{12}}{48}\right)\tilde{\alpha_2}^2 } \nonumber \\&& - \frac{800 \tilde{\alpha_2}^2 (ar-m) }{3}.
\end{eqnarray}
\begin{figure*}

\begin{tabular}{|c|c|c|c|}
\hline
\includegraphics[width= 8 cm, height= 5 cm]{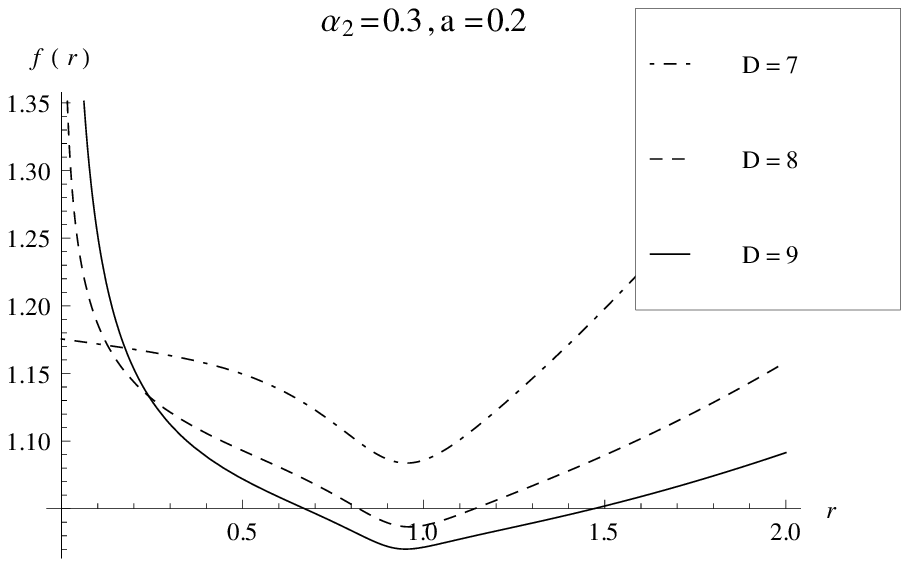}&
\includegraphics[width= 8 cm, height= 5 cm]{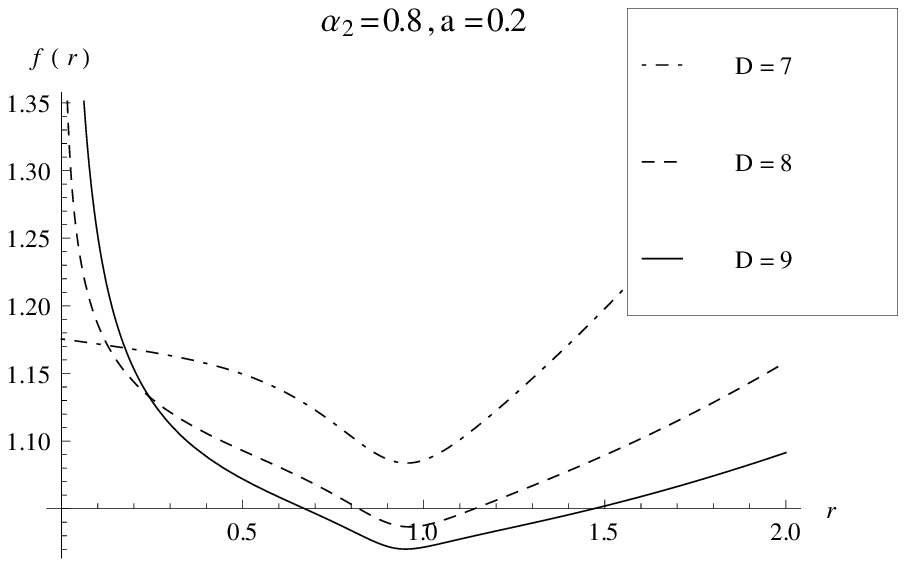}
\\
\hline
\includegraphics[width= 8 cm, height= 5 cm]{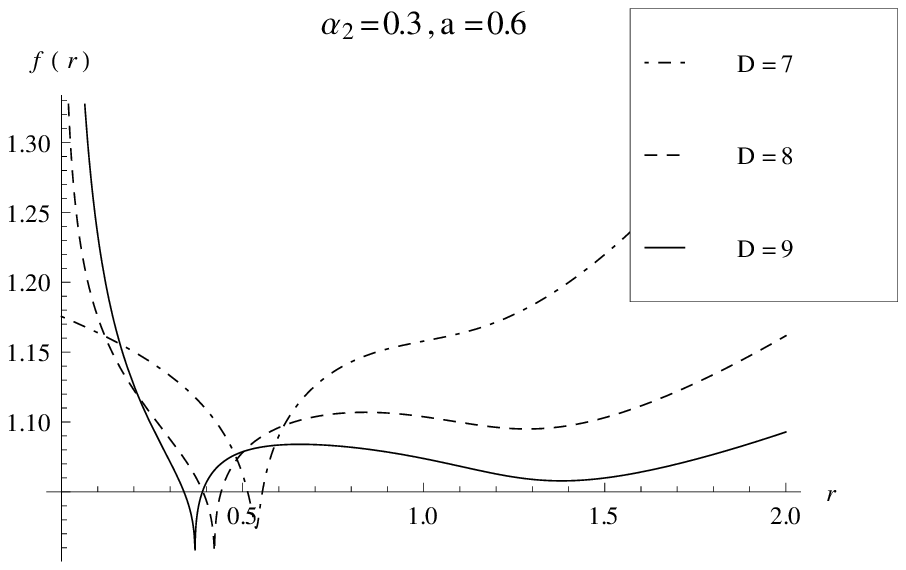}&
\includegraphics[width= 8 cm, height= 5 cm]{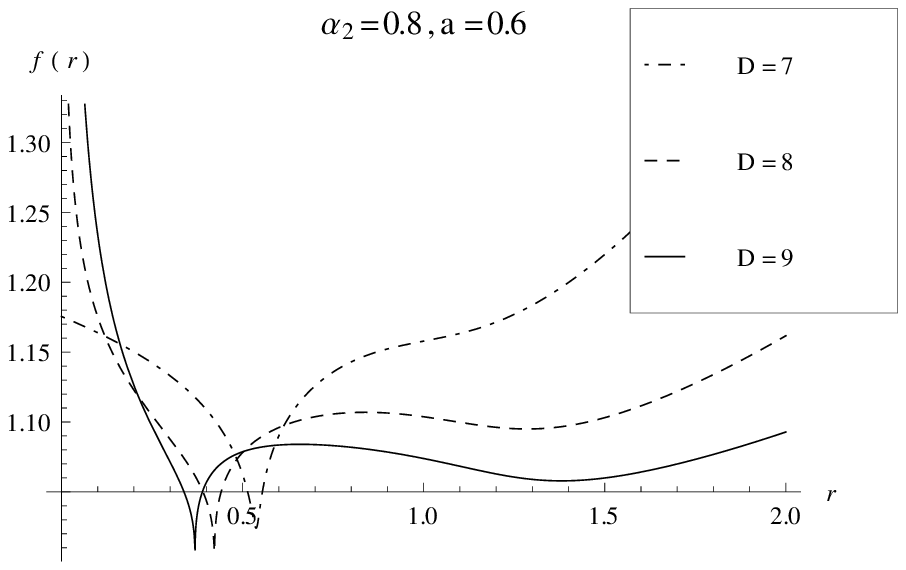}
\\
\hline
\end{tabular}
\caption{\label{ELLf(r)}  Plot of metric function $f(r)$ vs $r$ for the ELL case with $\alpha_3=1$ and $M=1$ in various dimensions. (Left): $\alpha_2=0.3$ and $a=0.2$  and $0.6$ (top to bottom). (Right): $\alpha_2=0.8$ and $a=0.2$ and $0.6$ (top to bottom). }
\end{figure*}
\begin{figure*}

\begin{tabular}{|c|c|c|c|}
\hline
\includegraphics[width= 8 cm, height= 5 cm]{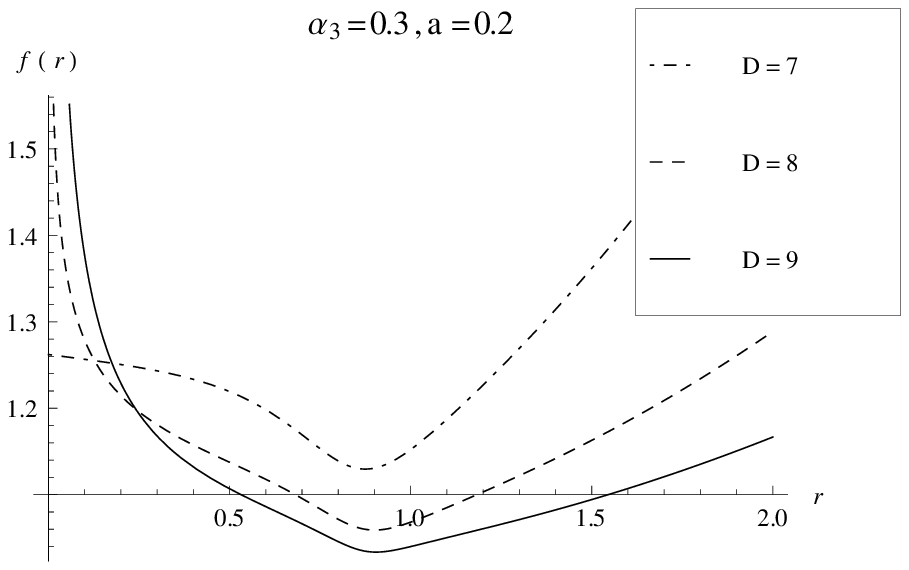}&
\includegraphics[width= 8 cm, height= 5 cm]{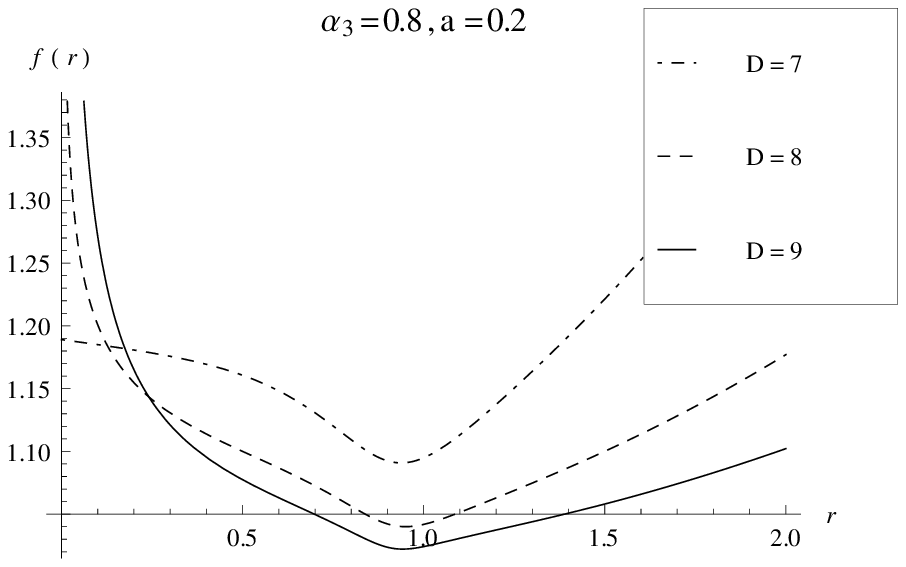}
\\
\hline
\includegraphics[width= 8 cm, height= 5 cm]{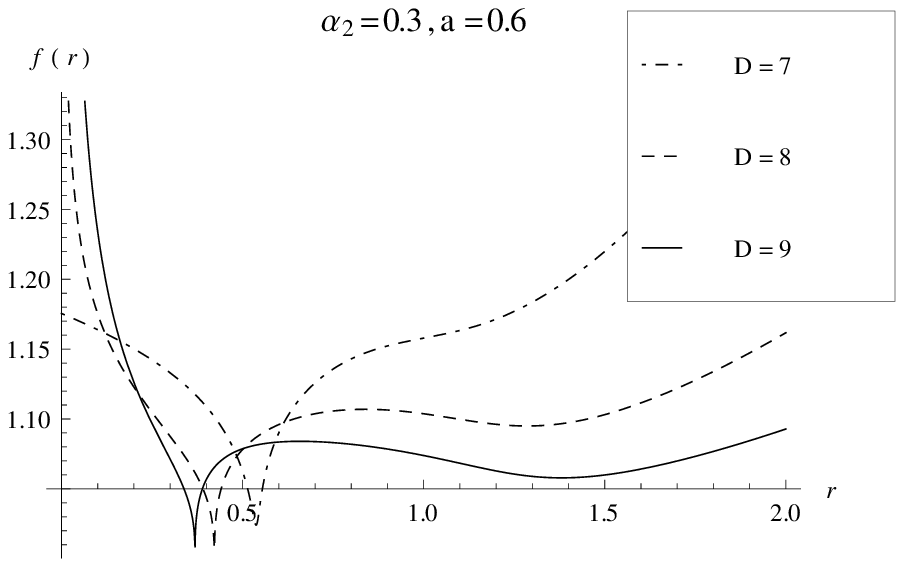}&
\includegraphics[width= 8 cm, height= 5 cm]{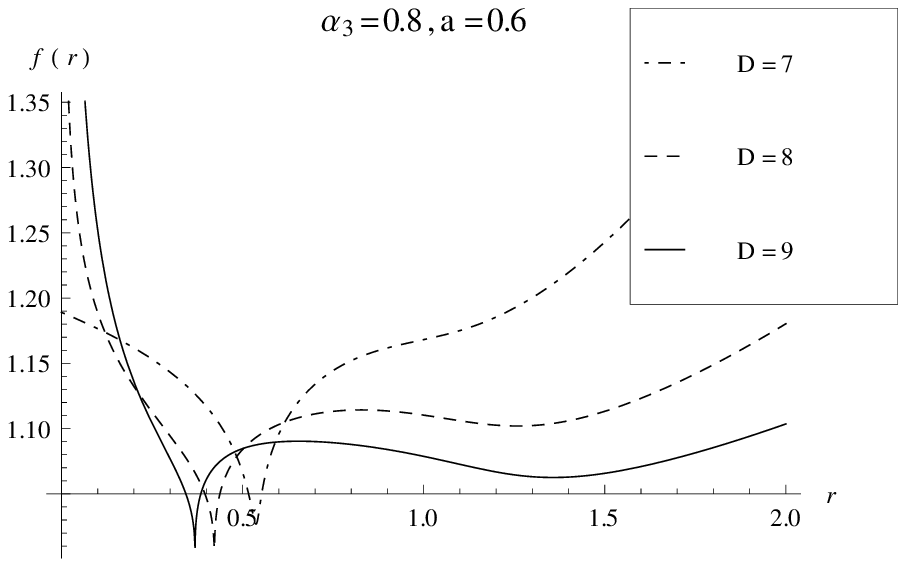}
\\
\hline
\end{tabular}
\caption{\label{ELLf(r)1}  Plot of metric function $f(r)$ vs $r$ for the ELL  case with $\alpha_2=1$ and $M=1$ in various dimensions. (Left): $\alpha_3=0.3$ and $a=0.2$  and $0.6$ (top to bottom). (Right): $\alpha_3=0.8$ and $a=0.2$ and $0.6$ (top to bottom). }
\end{figure*}

\subsection{Einstein-Lovelock (ELL) case}
In $D$-dimensional spacetime, we can see the roles of both second- and third-order Lovelock parameters simultaneously. A simplifying, yet
interesting case in the solution (\ref{sol:ll}) can be obtained if
$\tilde{\alpha}_{3}=\frac{\tilde{\alpha}_{2}^{2}}{3}.$ The metric function
$f(r)$ reads simply%
\begin{eqnarray}
f(r)&=&1+ \epsilon_3(r) \Delta^{1/3} -\frac{ \epsilon_4(r) }{\Delta^{1/3}},
\end{eqnarray}
with
\begin{eqnarray}
\epsilon_3(r) & =& \frac{1}{3^{2/3}n \tilde{\alpha}_3 r^n },\; \epsilon_4(r)=\frac{n r^{n+4} }{3^{1/3}}, \nonumber \\
\Delta &= & (9 r^5 (ar - m) + \sqrt{3}\sqrt{\delta} )\tilde{\alpha}_3^2 n^2  r^{2n}, \nonumber \\
\delta &= & \frac{n^2 r^{2n+12}+27 \tilde{\alpha}_3 (r a - m)^2 r^{10}}{\tilde{\alpha}_3}.
\end{eqnarray}
In Figs.~\ref{ELLf(r)} and \ref{ELLf(r)1}, we have plotted $f(r)$ for the ELL case in various dimensions. Figure~\ref{ELLf(r)} shows the variation of $f(r)$ taking fixed $\alpha_2$ and varying $\alpha_3$; Fig.~\ref{ELLf(r)1} shows the variation of $f(r)$ taking fixed $\alpha_3$ and varying $\alpha_2$.

The above solution in seven dimensions simplifies to
\begin{eqnarray}
f(r)&=& 1+ \epsilon_5(r) \Delta^{1/3} -\frac{ \epsilon_6(r) }{\Delta^{1/3}},
\end{eqnarray}with
\begin{eqnarray}
\epsilon_3(r) & =& \frac{3^{1/3} (25)^{1/3}}{15 \tilde{\alpha}_3 r^5 },\;
\epsilon_4(r) = \frac{r^9 3^{2/3 }(25)^{2/3}}{15}, \nonumber \\
\Delta &=&( 9 r^5 (ar - m) + \sqrt{3} \sqrt{\delta} )\tilde{\alpha}_3^2 r^{10}, \nonumber \\
\delta &= & \frac{25 r^{22}+27 \tilde{\alpha}_3 (r a - m)^2 r^{10}}{\tilde{\alpha}_3}.
\end{eqnarray}

\end{document}